\documentclass[apj, revtex4-1]{emulateapj}

\usepackage{epsf}
\usepackage{color}
\usepackage{amsmath}
\usepackage{graphicx}
\usepackage[colorlinks,urlcolor=magenta,citecolor=blue,linkcolor=blue]{hyperref}
\usepackage{natbib}
\usepackage{etoolbox}
\usepackage{threeparttable}

\graphicspath{{./figures/}}

\newcommand{\planck}{\textit{Planck}}

\newcommand{\sdssg}{\hbox{$g$}}
\newcommand{\sdssr}{\hbox{$r$}}
\newcommand{\sdssi}{\hbox{$i$}}
\newcommand{\sdssz}{\hbox{$z$}}

\newcommand{\lstar}{\hbox{L$_{\star}$}}

\newcommand{\msol}{\hbox{$\mathrm{M}_\odot$}}

\newcommand{\kms}{\hbox{km~s$^{-1}$}}

\newcommand{\degsq}{\hbox{degree$^2$}}
\newcommand{\arcminsq}{\hbox{arcmin$^2$}}

\newcommand{\perarcminsq}{\hbox{arcmin$^{-2}$}}

\newcommand{\permpc}{\hbox{Mpc$^{-1}$}}

\newcommand{\perpixel}{\hbox{pixel$^{-1}$}}

\newcommand{\eg}{e.g.}
\newcommand{\ie}{i.e.}

\newcommand{\citeeg}[1]{(\eg, \citealt{#1})}
\newcommand{\ignore}[1]{}

\makeatletter
\patchcmd{\NAT@citex}
{\@citea\NAT@hyper@{%
		\NAT@nmfmt{\NAT@nm}%
		\hyper@natlinkbreak{\NAT@aysep\NAT@spacechar}{\@citeb\@extra@b@citeb}%
		\NAT@date}}
{\@citea\NAT@nmfmt{\NAT@nm}%
	\NAT@aysep\NAT@spacechar\NAT@hyper@{\NAT@date}}{}{}

\patchcmd{\NAT@citex}
{\@citea\NAT@hyper@{%
		\NAT@nmfmt{\NAT@nm}%
		\hyper@natlinkbreak{\NAT@spacechar\NAT@@open\if*#1*\else#1\NAT@spacechar\fi}%
		{\@citeb\@extra@b@citeb}%
		\NAT@date}}
{\@citea\NAT@nmfmt{\NAT@nm}%
	\NAT@spacechar\NAT@@open\if*#1*\else#1\NAT@spacechar\fi\NAT@hyper@{\NAT@date}}

\shorttitle{}
\shortauthors{BOADA ET AL.}

\slugcomment{ \it Accepted for Publication in the Astrophysical Journal}

\begin{document}

\title{High Confidence Optical Confirmations Among the High Signal-to-Noise \planck\ Cluster Candidates}

\author{\sc Steven Boada\altaffilmark{1},
John P. Hughes\altaffilmark{1},
Felipe Menanteau\altaffilmark{2,3},
Peter Doze\altaffilmark{1},
L. Felipe Barrientos\altaffilmark{4},
L. Infante\altaffilmark{4}
}

\altaffiltext{1}{Department of Physics and Astronomy, Rutgers, the State University of New Jersey, 136 Frelinghuysen Road, Piscataway, NJ 08854-8019, USA; \href{mailto:boada@physics.rutgers.edu}{boada@physics.rutgers.edu}}
\altaffiltext{2}{National Center for Supercomputing Applications, 1205 West Clark St., Urbana, IL 61801, USA}
\altaffiltext{3}{Department of Astronomy, University of Illinois at Urbana-Champaign, 1002 W. Green Street, Urbana, IL 61801, USA}
\altaffiltext{4}{Instituto de Astrof\'isica y Centro de Astroingenie\'ia, Facultad de F\'isica, Pontificia Universidad Cat\'olica de Chile, Casilla 306, Santiago 22, Chile}

\begin{abstract}
\noindent We report on newly identified galaxy clusters from the high signal-to-noise ($>5$-$\sigma$) end of the second all-sky \planck\ Sunyaev-Zel'dovich (SZ) catalog (PSZ2). The clusters are identified in deep, optical imaging from the Kitt Peak National Observatory 4m Mayall telescope taken between 2014 and 2017. Here we focus on the highest richness systems, and identify galaxy clusters through a combination of the maxBCG algorithm and visual image inspection. Galaxy clusters are considered to be confirmed if they are both rich and spatially coincident ($\lesssim 6'$) with the reported PSZ2 position. Of the 85 fields containing unconfirmed PSZ2 candidates observed, we find 15 (17.6\% of the observed sample) corresponding galaxy clusters ($ 0.13 < z < 0.78$), 12 of which are previously unrecognized as counterparts. To explain this low identification fraction, we consider three possible scenarios: that clusters are (1) mostly at low-z, (2) mostly at high-z, or (3) located in fields with high object density.  None of these scenarios alone can account for the low purity of rich galaxy clusters among the high signal-to-noise PSZ2 unconfirmed candidates.
\end{abstract}

\section{Introduction}\label{sec:intro}
Galaxy clusters, especially massive clusters, are extraordinary objects that contain vital clues to the structure and evolution of the universe. The generally accepted $\Lambda$CDM cosmological model makes detailed predictions about the number and total mass distribution of galaxy clusters throughout the universe. These galaxy clusters, particularly at high redshift, play a crucial role in constraining large scale structure formation and evolution from measurements of the overall matter density and the amplitude of matter fluctuations \citeeg{Evrard1989,Henry1991,Borgani2001,White1993,White1993b,Eke1996,Donahue1998}, the dark energy equation of state \citeeg{Henry2004,Mantz2008,Vikhlinin2009a}, and the potential falsification of $\Lambda$CDM from the existence of extreme clusters \citeeg{Mortonson2011, Harrison2012, Harrison2013, Waizmann2013}.

Large-area sky surveys, both currently underway and planned, are identifying many tens of thousands of galaxy clusters for both detailed examinations of our cosmological models and as probes of fundamental physics. Clusters have been identified in numerous ways: as overdensities of galaxies in optical surveys \citeeg{Abell1958,Postman1996}, as extended X-ray sources \citeeg{Gioia1990,Ebeling1998,Boehringer2000}, as clusters of mass in gravitational-weak-lensing surveys \citeeg{Wittman2006}, and most recently as ``shadows'' or ``holes'' in the Cosmic Microwave Background (CMB).  Today, large number of galaxy clusters are being identified in ground based, millimeter wave surveys conducted by the Atacama Cosmology Telescope (ACT; \citealt{Swetz2011}) and the South Pole Telescope (SPT; \citealt{Carlstrom2011}), among others. These surveys have identified hundreds of massive galaxy clusters up to redshifts of $z \sim 1.4$ \citeeg{Vanderlinde2010, Menanteau2010, Marriage2011, Hasselfield2013a, Bleem2015, Hilton2018}. This detection approach exploits the cluster's hot intracluster medium which imprints a telltale signature on the Cosmic Microwave Background (CMB) through the Sunyaev-Zel'dovich effect (SZ; \citealt{Sunyaev1972}) effect.

Deep, wide field photometric surveys such as the ongoing Hyper Suprime-Cam Survey \citep{Aihara2018a} and Dark Energy Survey (DES; \citealt{DES2005}), as well as future surveys like the Large Synoptic Survey Telescope (LSST; \citealt{LSST2012}), \textit{Euclid}, and WFIRST will discover many thousands of additional galaxy clusters at redshifts to $z=1$ \citeeg{DarkEnergySurveyCollaboration2016} and beyond. While a powerful discovery method, photometric surveys suffer from cosmological surface brightness dimming that reduces their sensitivity at high redshift \citeeg{Calvi2014}.

Using the SZ effect to discover clusters of galaxies has the distinct advantage that the surface brightness of the SZ effect does not dim with increasing redshift. This allows fairly homogeneous samples of massive clusters to be detected out to arbitrary distances. Now, \planck\ \citep{Tauber2010, PlanckCollaboration2011} has released an all-sky SZ sample (hereafter PSZ; \citealt{PlanckCollaboration2014, PlanckCollaboration2015a}) that contains 1943 detections with 1330 confirmed clusters and another 613 (as of PSZ catalogs\footnote{\url{http://szcluster-db.ias.u-psud.fr/sitools/client-user/SZCLUSTER_DATABASE/project-index.html}} versions 2.1 and 1) unconfirmed cluster candidates. Clusters were initially confirmed by cross correlating with previous catalogs (see Section 4; \citealt{PlanckCollaboration2014}). More recently, dedicated follow up of still-unconfirmed clusters has begun in earnest \citeeg{Liu2015a, PlanckCollaboration2015, PlanckCollaboration2016a, VanderBurg2016, Burenin2017, Barrena2018, Amodeo2018, Streblyanska2018}.

This study was motivated by the release of the first PSZ all-sky catalog in 2013. We originally posited that the vast majority of the candidates must lie at $z>0.4$ because the \planck\ confirmation process \citep{PlanckCollaboration2014} mostly relied on existing catalogs which have a preference for low-$z$ clusters. Furthermore, the confirmed sample of the PSZ2 catalog has only a small fraction (3\%) of $z > 0.6$ clusters compared to that expected ($\sim20$
\%) based on the theoretical halo mass function \citeeg{Jenkins2001, Tinker2008} for mass limit of $6\times10^{14}$ \msol. Of particular interest to us is the search for the most massive clusters at high redshift, such as ACT-CL J0102$-$4915 (hereafter ``El Gordo''), which weighs in at $\sim2 \times 10^{15}$ \msol\ at a redshift of $z=0.870$ \citep{Menanteau2012}. The Tinker halo mass function suggests as many as four clusters as massive as this at $z>0.6$ in the full sky area covered by the \planck\ PSZ catalog (83.7\% of the sky). If other clusters as massive as ``El Gordo'' exist, they are hiding as high-significance candidates within the objects in this \textit{all-sky} catalog. In the discussion section below we examine this argument more carefully.

In this paper we report on our first efforts to classify unconfirmed PSZ cluster candidates using optical observations. This paper is organized as follows: sections \ref{sec:design} through \ref{sec:analysis} describe the design, observations, data reduction and calibration, and creation of derived data products. In Section~\ref{sec:results}, we present the main results of our observations, and discuss those results in Section~\ref{sec:discussion}. In Section~\ref{sec:summary}, we summarize the key results and conclude.

Unless otherwise noted, throughout this paper, we use a concordance cosmological model ($\Omega_\Lambda = 0.7$, $\Omega_m = 0.3$, and $H_0= 70$ \kms \permpc), assume a Chabrier initial mass function \citep{Chabrier2003}, use AB magnitudes \citep{Oke1974} and quote uncertainties at the 1-$\sigma$ level.

\section{Observational Strategy}\label{sec:design}

The core of our observational design relies on the use of optical imaging to confirm the SZ detections as real clusters and provide photometric redshifts using the multi-color information. This design is based on the previous success with the ACT cluster confirmation process using 4-m class telescopes. Although \planck's larger beam size (compared to both ACT and SPT) makes it more sensitive to clusters at lower redshifts (due to their larger projected area on the sky), among the confirmed clusters in the recently released all-sky \planck\ SZ catalog are the two highest significance high-redshift SZ detections from ACT (as well as several other ACT and SPT clusters).

Our strategy for this project is to use the Kitt Peak National Observatory (KPNO) Mayall-4m telescope as the first and fundamental step to confirm the highest significance detections in the PSZ2 catalog that are visible across the entire northern sky. Optical imaging should be sufficient to confirm nearly all of the candidates. However, for the highest redshift candidates, near-IR imaging will be necessary. Those candidates with some evidence for a high-$z$ brightest cluster galaxy (BCG) should be targeted with near-IR observations to confirm the presence of the BCG and to establish a red sequence of cluster members.

Following closely the procedure used for ACT follow-up \citeeg{Menanteau2013}, targets are prioritized by SZ SNR. We choose to initially report on targets with PSZ2 SNR $>5$ as the statistical reliability of PSZ2 cluster candidates should be quite high: according to the \planck\ team $\sim90$\% of candidates at SNR $>5$ are expected be ``real'' clusters (see Figure 11; \citealt{PlanckCollaboration2015a}).

\subsection{Observations}\label{sec: observations}
All observations were conducted with the KPNO Mayall telescope. The optical observations were made with the MOSAIC camera mounted at the prime focus. Two detector packages were used for the observations. The earlier MOSAIC1.1 instrument consisted of eight $2048\times4096$ SITe CCDs, arranged $2\times4$, separated by a gap $∼\sim50$ pixels wide with a pixel scale of $0\farcs26$ \perpixel. MOSAIC1.1 was replaced with MOSAIC3, in mid-2015, and consists of four new 4k$\times$4k, 15 micron pixel, 500-micron thick LBNL deep-depletion CCDs. Because the only change from MOSAIC1.1 to MOSAIC3 are the CCDs and controllers, both versions have a $36' \times 36'$ field-of-view.

The optical observing strategy consists of targeted \sdssg\sdssr\sdssi\sdssz\ observations of individual candidates with total exposure times of 360 s, 360 s, 1100 s and 1100 s. The final exposures consist of four dithered positions with individual exposures of 90 s for the \sdssg\sdssr-bands or 275 s for the \sdssi\sdssz-bands. These exposure times are designed to ensure the unambiguous detection of the faint galaxies in the red cluster sequence up to $z \sim 1.0$ and of brightest cluster galaxies (BCGs) to higher redshifts. The choice of filters in our program is driven by the need to segregate early-type galaxies in the cluster through their colors (or photometric redshifts) by sampling blue-ward and red-ward of the 4000\AA\ break over a broad range of redshifts.

\section{Data Reduction and Calibration}\label{sec:data reduction}
Standard image reductions including subtraction of dark frames, flat fielding, sky-subtraction, and bad pixel masking were performed by the National Optical Astronomy Observatory virtual observatory (VO) using the MOSAIC \citep{Valdes2007} science pipelines. The resultant FITS files consist of fully reduced images with either all single exposure CCDs mosaicked into a single image extension (as in the case of Mosaic1.1) or as a multi-extension FITS file with each single exposure CCD occupying a separate extension.

We then mosaic each separate exposure into a master mosaic as described in the following section.

\subsection{Mosaicking}\label{sec:mosaicks}
Combined mosaics are created with \textsc{swarp} (version 2.38.0; \citealt{Bertin2002}). We create three distinct types of mosaics. The individual dither frames are stacked and then median combined to produce the final completed science mosaic. A ``detection'' is created by combining select science mosaics into a ``chi2'' image using a combination of the \sdssi- and \sdssz-band, when both are available and of sufficient quality, or the \sdssi-band when only one is present. Finally, we create a set of mosaics to produce the three color image used for cluster finding. We median combine the \sdssg\sdssr\sdssi\sdssz\ science mosaics into a ``blue'' (\sdssg-band), ``green'' (\sdssr-band), and ``red'' (\sdssi\sdssz-band) mosaic. All final mosaics have a pixel scale of $0\farcs25$ \perpixel. The final exposure time is calculated as the median exposure time of the combined images, and similarly the final air mass is median of the individual air masses. We also compute the final seeing for our mosaics by examining the mosaics with the IRAF \citep{Tody1993} task IMEXAMINE. Typical seeing is approximately $1\farcs2$ in the \sdssi-band.

Because of the different detector orientations, final science mosaics from images taken with Mosaic1.1 and Mosaic3 differ in size. Science mosaics are typically $\sim0.45$ \degsq\ for observations taken with the Mosaic1.1 and $\sim1.0$ \degsq\ for Mosaic3 observations.

\subsection{Source Extraction and Photometry}\label{sec:sextractor}
For source extraction and photometry estimation we use Source Extractor\footnote{\url{https://github.com/astromatic/sextractor}} (hereafter SExtractor; version $2.19.5$; \citealt{Bertin1996}) run in dual image mode with the ``chi2'' detection image as the detection image. See Section~\ref{sec:mosaicks}. We configure SExtractor to detect objects with a minimum threshold of 1.5$\sigma$ and a minimum area of 12 pixels.
Our pipeline reports source photometry in magnitudes derived from summing the flux within the isophotal area reported by SExtractor. When the isophotal flux is flagged by SExtractor as unreliable, we fall back to magnitudes reported by SExtractor as \textsc{MAG\_AUTO} with a Kron factor \citep{Kron1980} of $2.5$ and a minimum radius of 3.5 pixels ($0\farcs88$).

Because our images are spread across the entire northern sky, we have implemented dust corrections for every source across all four bands. This correction is particularly important in ensuring accurate galaxy colors and correct photometric redshifts. We utilize the infrared maps and correction routines provided by \cite{Schlegel1998} for this dust correction.

\subsection{Astrometric Calibration}
Each of the final science mosaics produced in the previous section is first astrometrically aligned with \textit{Gaia} \citep{GaiaCollaboration2016} Data Release 1 \citep{GaiaCollaboration2016a} using \textsc{scamp} \citep{Bertin2006} as a part of \textsc{photometrypipeline}\footnote{\url{https://github.com/mommermi/photometrypipeline}} (PP; \citealt{Mommert2017}).

Sources are extracted from the mosaics with a SNR of at least 10 and with a minimum area of at least 12 pixels. The extracted sources are then matched to the \textit{Gaia} data and a new astrometric solution is calculated. Matching all science images to the \textit{Gaia} world coordinate system ensures that we have a common alignment across all observing runs. Because the initial astrometric solution from the VO is quite accurate, the
typical root mean squared (RMS) errors reported by \textsc{scamp} are less than $0\farcs05$.

\subsection{Photometric Calibration}
After the mosaics have been astrometrically aligned, we use PP to produce a photometric solution. PP calculates a photometric zero-point in each of our observed bands by comparing field stars located throughout the science mosaic to known photometry from large-area sky surveys. Because our sources are spread across the entire northern sky, and because we prefer to minimize the number of differences between photometric solutions we are limited to two optical surveys. We first seek photometric data from the \textit{Sloan Digital Sky Survey} (SDSS; \citealt{York2000}) Data Release 13 (DR13; \citealt{Albareti2017}). When our target does not lie within the SDSS footprint we utilize the Panoramic Survey Telescope and Rapid Response System (Pan-STARRS; \citealt{Chambers2016}) Data Release 1 (hereafter PS1; \citealt{Flewelling2016}). Both surveys provide accurate \sdssg\sdssr\sdssi\sdssz\ magnitudes and large on-line queriable databases for rapid automated calibration.

Sources are extracted from the combined mosaics with a $3''$ (12 pixel) diameter aperture; stars in our science images with SNR $\ge10$ are matched to a survey catalog and a photometric zero-point is determined. Stars selected from the large-area sky surveys have either \sdssr-band (for PS1) or \sdssg-band (for SDSS) magnitudes $<21$ and accurate catalog photometry (\eg, the ``clean'' SDSS flag). We use at least 50\% of the available stars to derive the zero-point resulting in zero-points calculated from approximately $10-500$ stars and with typical uncertainties of $0.05$ mag for the \sdssg\sdssr\sdssi-bands and $0.16$ mag for the \sdssz-band.

\section{Analysis}\label{sec:analysis}
\subsection{Photometric Redshifts}
We determine photometric redshifts (photo-$z$) from the four-band optical imaging catalogs using the Bayesian Photometric Redshifts (BPZ; \citealt{Benitez2000, Coe2006}) code following the same procedure as in \cite{Menanteau2009a}.

We assess the effectiveness of our photo-$z$ estimates by comparing with the available spectroscopic redshifts (spec-$z$) from the SDSS. We use three diagnostics to gauge photo-$z$ accuracy. First, we report the full scatter between the photo-$z$ and spec-$z$, defined as:
\begin{equation}\label{eqn:scatter}
	\sigma_f = \mathrm{RMS}[\delta z/(1+z_\mathrm{spec})]
\end{equation}
where $\delta z = z_\mathrm{spec} - z_\mathrm{phot}$. Second, we report the normalized median absolute deviation (NMAD; \citealt{Ilbert2009, Dahlen2013, Molino2017}), given as
\begin{equation}\label{eqn:nmad}
	\sigma_\mathrm{NMAD} = 1.48 \times \mathrm{median} \bigg{(} \frac{|\delta z|}{1+z_\mathrm{spec}} \bigg{)}.
\end{equation}
which provides an estimate of the scatter resistant to catastrophic outliers. Finally, the catastrophic outlier fraction (OLF) where we define a catastrophic outlier (following \citealt{Molino2017}) as,
\begin{equation}\label{eqn:OLF}
	\eta = \frac{|\delta z|}{(1+z_\mathrm{spec})} > 5 \times \sigma_\mathrm{NMAD}.
\end{equation}

\begin{figure}
	\includegraphics[width=\columnwidth]{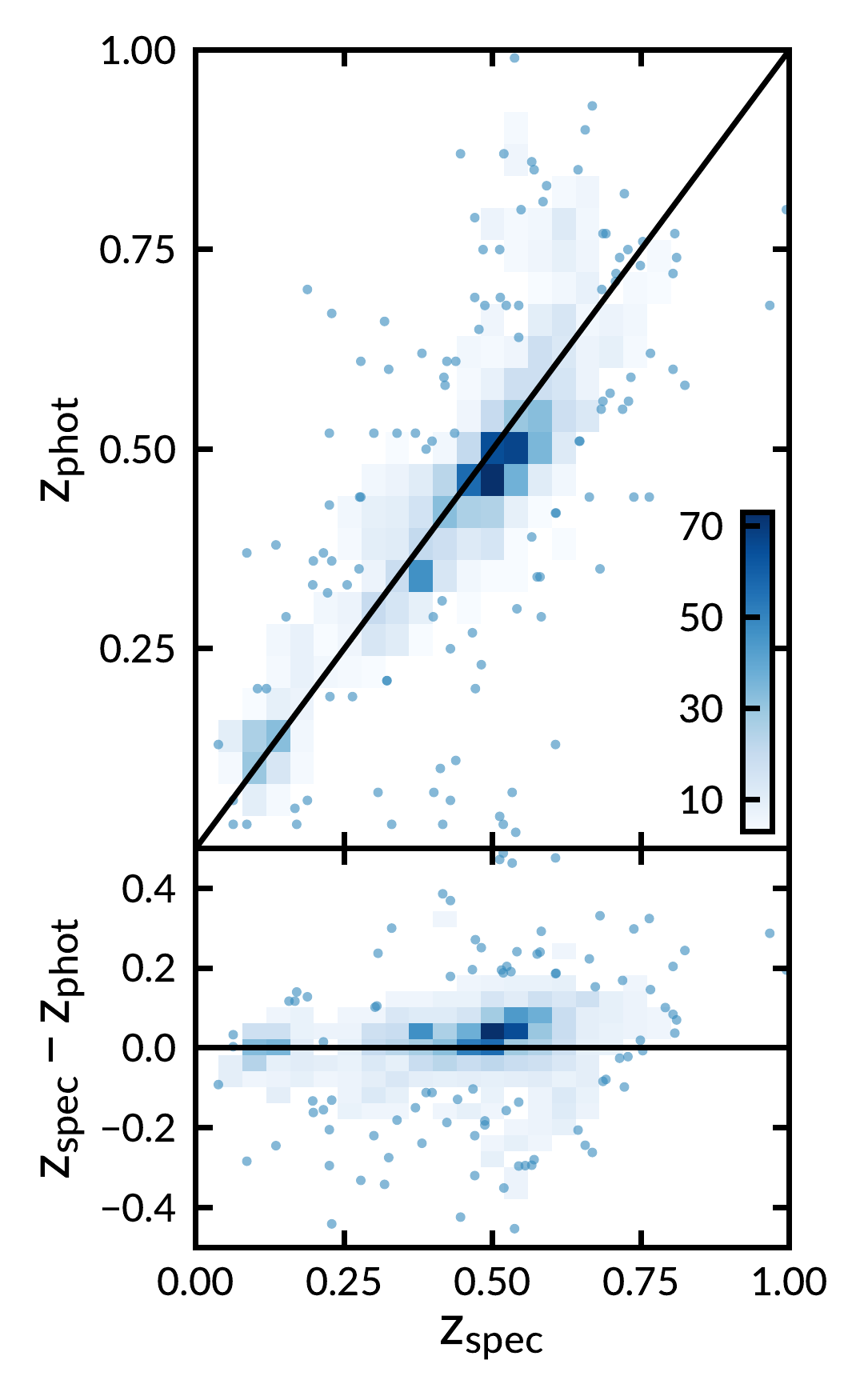}
	\caption{Comparison between photometric and spectroscopic redshifts for 1588 elliptical galaxies with spectroscopic redshifts from the SDSS. \textit{Top:} The photometric redshifts are those reported by BPZ with a custom empirical prior on galaxy brightness for the photometric redshifts. \textit{Bottom:} The difference between the spectroscopic and photometric redshifts, $z_\mathrm{spec} - z_\mathrm{phot}$, as a function of spectroscopic redshift. In both panels the shading scales (darker is greater) with the density of points where bins with fewer than three points are shown as single points. The solid black line shows the 1:1 relation.}
	\label{fig:photozspecz}
\end{figure}

Figure~\ref{fig:photozspecz} shows the photo-$z$ performance as a function of the true spectroscopic redshift. Because we are primarily concerned with identifying clusters containing early-type galaxies, we show only galaxies classified E/S0 by BPZ. We find $\sigma_f = 0.067$, $\sigma_\mathrm{NMAD} = 0.048$, and an outlier fraction, $\eta = 0.9\%$.

\subsection{Recovery of the Brightest Cluster Galaxies}
We have designed our observations to detect BCGs to $z\sim1.5$. To quantify the actual depths of our images, we perform two distinct tests. First, we perform a Monte Carlo simulation by injecting artificial, elliptical galaxies into the science imaging and computing their recovery fractions. Second, we fit a power-law model to the bright side of each field's differential number count curve and identify the magnitude, $m$, at which  the observed differential, $dN/dm$, falls below the extrapolated power-law fit by a specified amount.

The Monte Carlo simulation broadly follows the procedure given in \cite{Menanteau2010a}. We create the artificial elliptical galaxies with the \textsc{modeling} package, part of \textsc{astropy} \citep{TheAstropyCollaboration2013}. The synthetic galaxies are created to have de Vaucouleurs \citep{DeVaucouleurs1948} profiles and surface brightnesses corresponding to their magnitude and assumed sizes. We inject the artificial galaxies into our science images with similar noise characteristics as their real counterparts.

We generate 10 rounds each spreading 100 elliptical galaxies randomly across our science imaging. Galaxies are placed at different random positions, from round to round, to suppress abnormally boosted recovery fractions due to source confusion. The artificial galaxies have total fluxes corresponding to apparent magnitudes between 20 mag $< \sdssi <$ 25 mag with $0.2$ mag spacing. We report the 80\% \sdssi-band limit for each field, which is the magnitude where 80\% of the simulated galaxies are recovered.

\begin{figure*}
	\centering
	\includegraphics[width=0.8\textwidth]{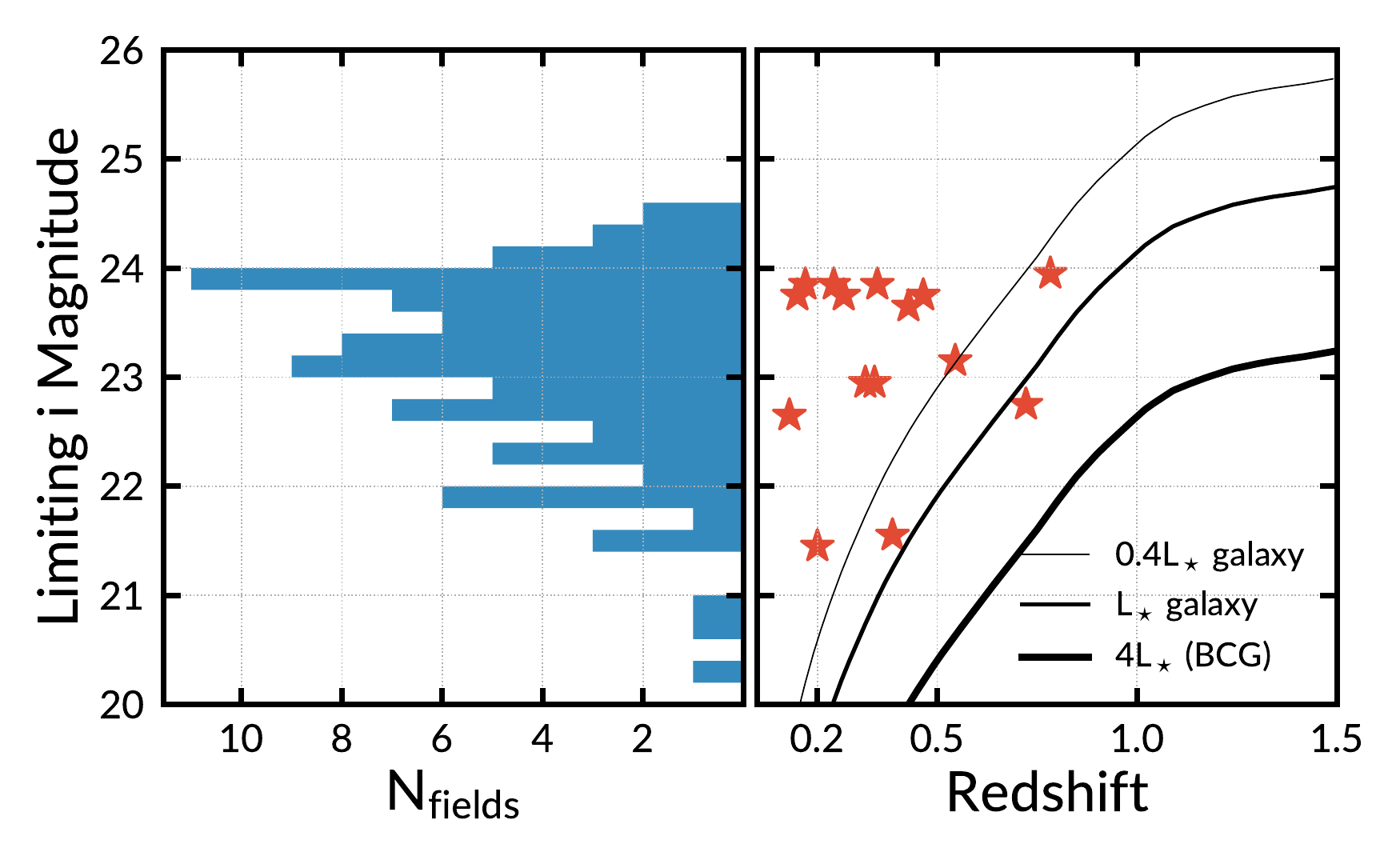}
	\caption{\textit{Left:} Histogram of the \sdssi-band magnitude corresponding to 80\% completeness in galaxy recovery. When 80\% completeness is not achieved we show the limiting magnitude with the highest completeness. \textit{Right:} Observed \sdssi-band magnitudes of $\lstar$, $0.4\lstar$, and $4\lstar$ (BCG) early-type galaxies as a function of redshift. We define an \lstar\ galaxy following \cite{Blanton2003} as a population of red galaxies at $z = 0.1$ and allow it to evolve passively. The orange stars indicate the field depths where we identify clusters and are plotted at the cluster's photometric redshift. The left and right panels can be combined to estimate the limiting redshift to which we could identify galaxy clusters.}
	\label{fig:recovery_redshift}
\end{figure*}

The second method to estimate the limiting magnitudes of our observations uses the calibrated SExtractor catalogs (see Section~\ref{sec:sextractor}). We bin the objects identified by SExtractor as galaxies in 0.5 mag wide bins from $15<$ \sdssi\ $<30$. We identify the peak of the differential number counts function ($dN/dm$ vs. $m$) in log space and fit a power-law model to the five magnitude bins on the bright side of the peak. The 80\% completeness limit is defined as the magnitude bin at which the observed counts are 80\% of the value from the extrapolated power-law.

The completeness limits reported by each method are consistent and provide an important check on one another as they are wholly independent estimates of the limiting \sdssi-band magnitude in each of our fields. We choose to focus on the limiting magnitudes reported by the number counts based method as it is less prone to errors from very bright stars or diffuse emission from the Milky Way also present in some fields.

With a limiting magnitude reported for each field, we are able to estimate the redshift to which we could reliably identify massive galaxy clusters. We compare the completeness limits of our observations to the expected (\ie, known) apparent magnitudes of galaxies in clusters as a function of redshift. We define the expected apparent \sdssi-band magnitude of a galaxy following the population of red galaxies defined by \cite{Blanton2003}. The 0.4\lstar, \lstar, and 4\lstar\ galaxies (representing faint members to bright BCGs) are composed of stars formed in a burst at $z = 5$, with solar metallicity \citep{Bruzual2003}, and allowed to evolve passively, $\tau = 1.0$ Gyr. We show the expected \sdssi-band apparent magnitude as a function of redshift for each galaxy in the right panel of Figure~\ref{fig:recovery_redshift}.

\subsection{Cluster Finding}\label{sec:cluster finding}
In this section, we briefly describe the algorithms and methods used to select the galaxy clusters from the multi-wavelength optical imaging. We follow the methods described in detail in \cite{Menanteau2009a, Menanteau2010} and direct the reader there for an in depth description and discussion of the methods. In short, we adopt a hybrid approach where we first manually identify (by eye) potential cluster BCGs and then use the maxBCG algorithm \citep{Koester2007} to identify cluster members.

We first create a three-color image using \textsc{stiff} (version 2.4.0; \citealt{Bertin2011}). The red, green, and blue channels are given by the corresponding combined mosaics described in Section~\ref{sec:mosaicks}. We then visually inspect an area of $6'$ in radius centered on the position of each unconfirmed PSZ detection. This search area should enclose the vast majority of clusters \citeeg{PlanckCollaboration2016a, Streblyanska2018} and allows for the inclusion of possibly disturbed cluster systems (\eg, see Figure 10; \citealt{PlanckCollaboration2015a}). This search radius, which is slightly larger than the expected positional uncertainty of \planck\ sources, will enable us to conduct a thorough search while still reporting high confidence confirmations. Potential BCGs are selected manually and are identified by their characteristic colors and accompanying member galaxies.

Once a potential BCG is selected, the maxBCG algorithm selects nearby member galaxies. We require these member galaxies to meet the following conditions: (1) they must be classified as either E or E/S0 type according to BPZ; (2)  they must have $|z_\mathrm{BCG} - z| < 0.05$ and be within a 1 Mpc projected radius of the BCG; (3) they must be fainter than the selected BCG and brighter than 0.4\lstar, where we have defined \lstar\ in the previous section; (4) they must have colors consistent with the cluster. To ensure the last point, we obtain a local color-magnitude relation for $\sdssg-\sdssr$ and $\sdssr-\sdssi$ and use a $3-\sigma$ mean sigma-clipping method to iteratively remove galaxies which have colors beyond 3-$\sigma$ from the local color-magnitude relationship. The potential member galaxies are iterated over until the number of cluster members converges.

The photo-$z$'s of the galaxies are combined using the same 3-$\sigma$ mean sigma-clipping algorithm to estimate the cluster's mean redshift, $z_\mathrm{cl}$. We use this cluster redshift measurement and the member selection criteria given previously to estimate the number of cluster members within 1 Mpc which we define as the richness \citeeg{Abell1958} of the cluster, Ngal. For three fields where we identify rich clusters, the depth of our imaging does not allow us to reliably detect galaxies to 0.4\lstar at the redshift of the cluster. These clusters can be easily identified among the orange star points in the right panel of Figure~\ref{fig:recovery_redshift}. The richness estimates for these three fields, should, therefore, be taken as lower limits. We note these lower limits in Table~\ref{tbl:results} and in our discussion of individual clusters in Section~\ref{sec:discussion}.

We correct the Ngal estimate by subtracting a statistical background of galaxies. We select background galaxies using the same redshift, luminosity, and color criteria as the member galaxies described previously. However, to ensure that we are not including galaxies belonging to the cluster itself, we only consider galaxies beyond 3 Mpc of each cluster's position. The number of background galaxies is scaled by the ratio of areas and then subtracted from the number of cluster members to provide a corrected Ngal, Ngal$_c$, which we then use to compute other important quantities. In practice the corrected number of galaxies is between 15\% and 20\% lower than the uncorrected number \citep{Menanteau2010}. We report Ngal$_\mathrm{c}$ for the remainder of this work.

\section{Results}\label{sec:results}
\begin{figure*}
	\centering
	\begin{tabular}{cc}
		\includegraphics[width=0.47\linewidth]{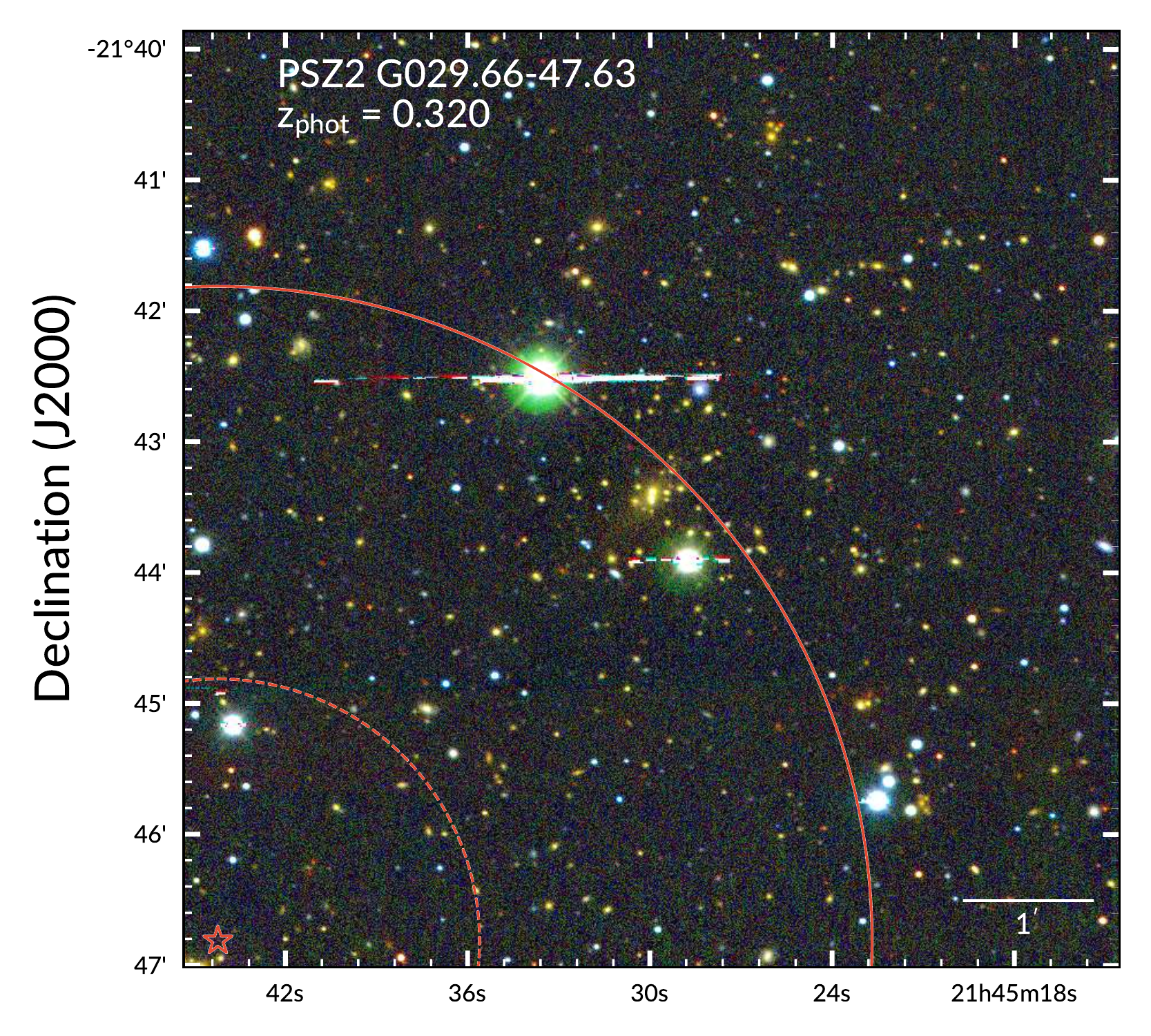}&
		\includegraphics[width=0.47\linewidth]{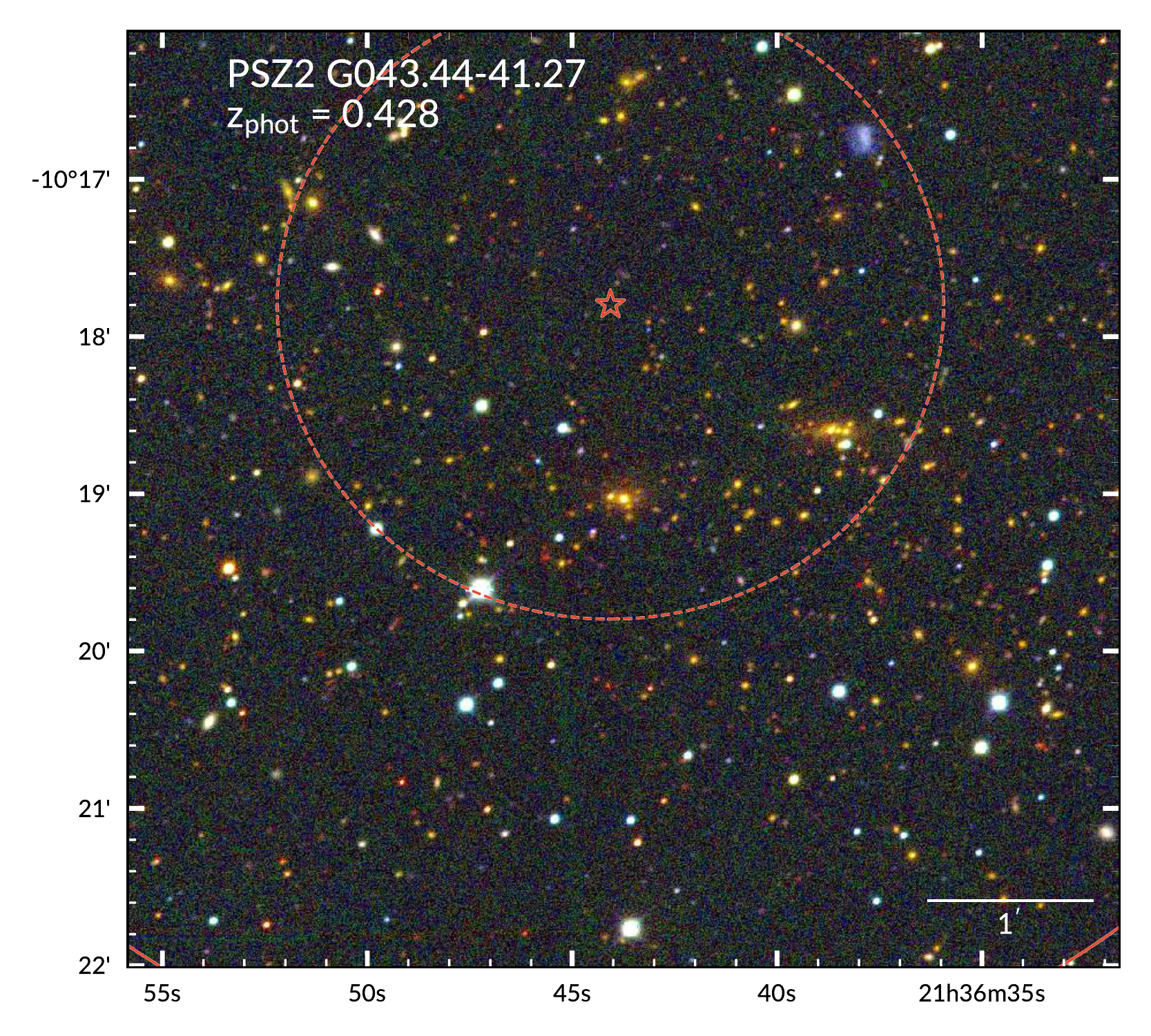}\\
		\includegraphics[width=0.47\linewidth]{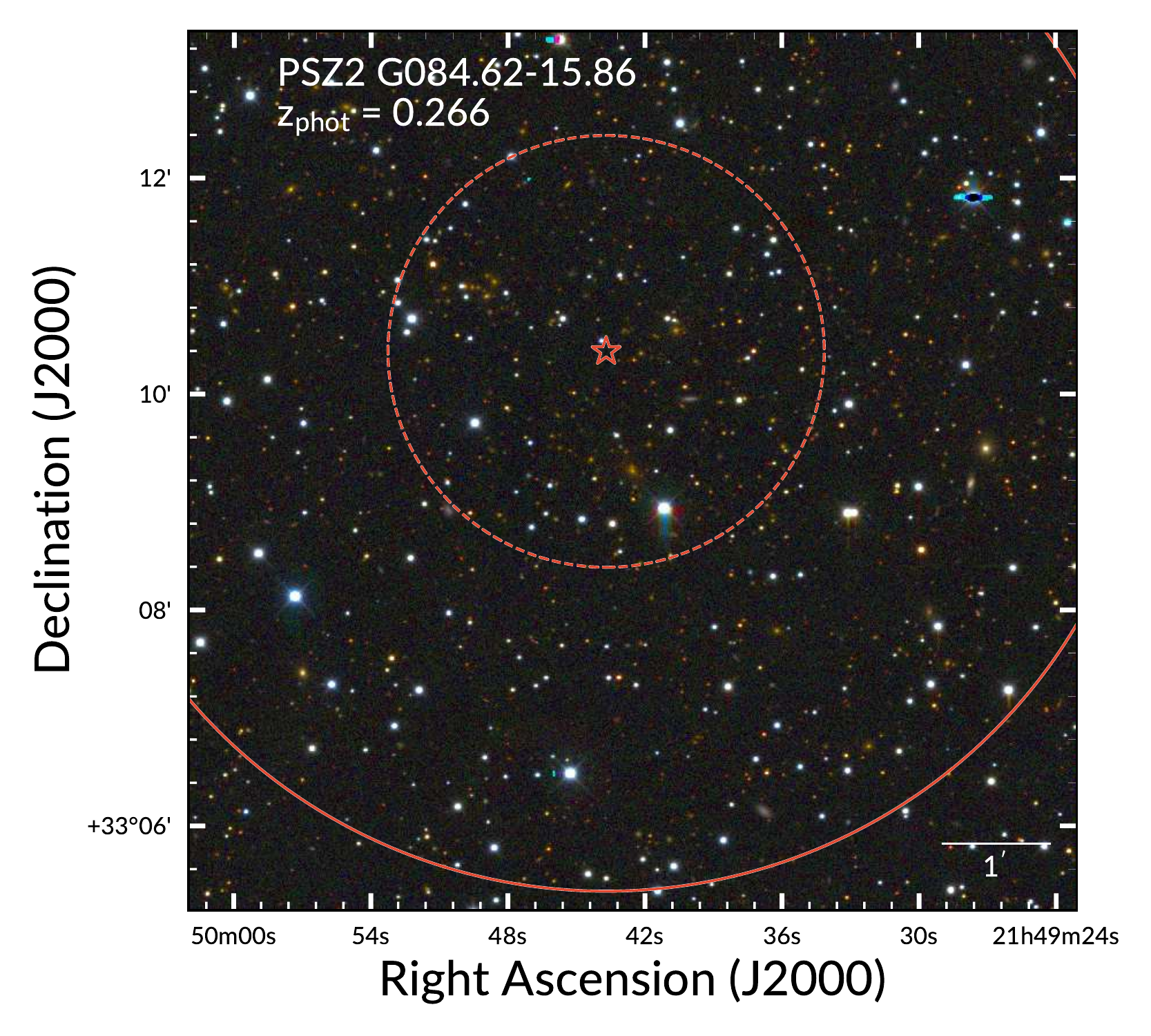}&
		\includegraphics[width=0.47\linewidth]{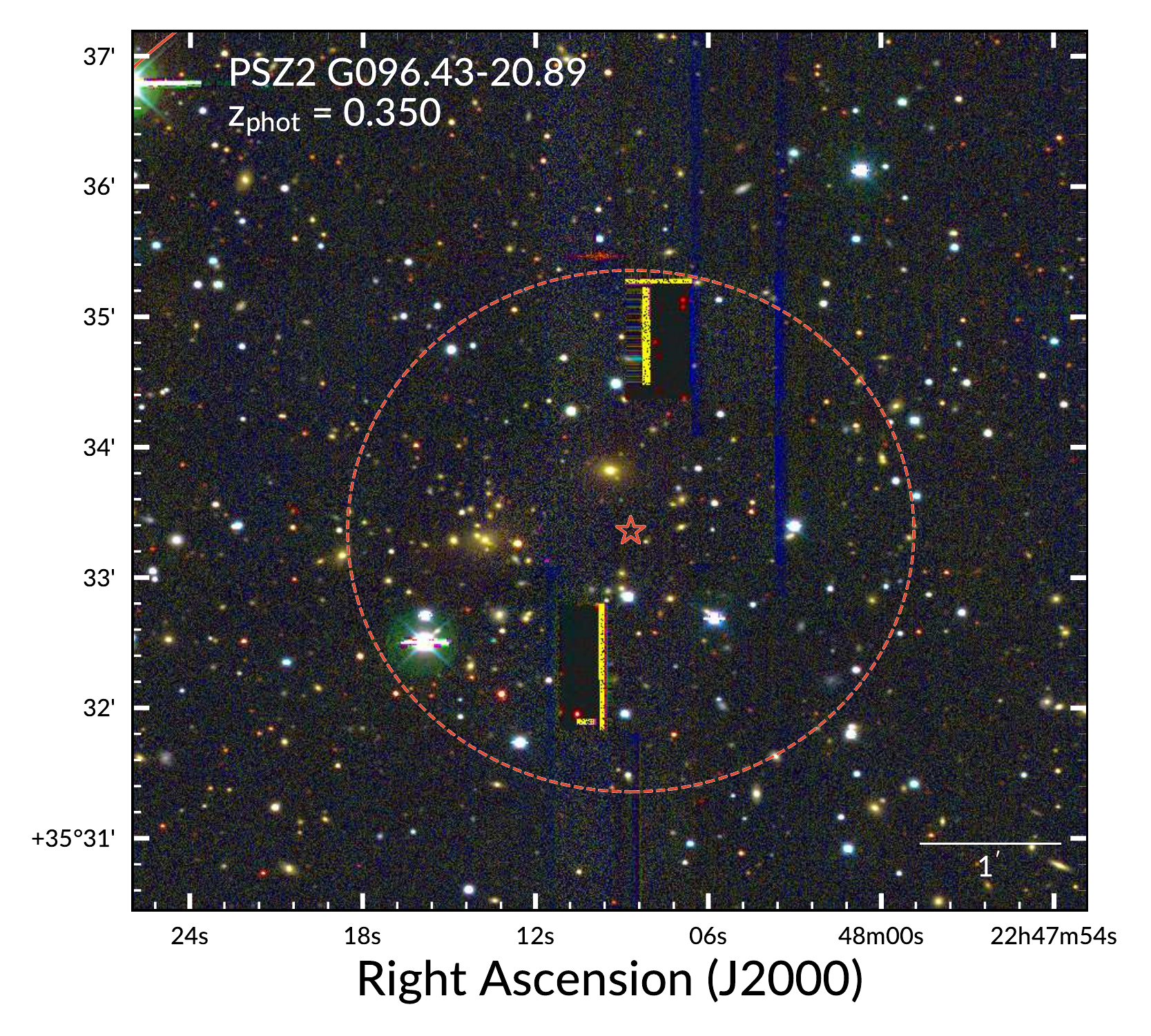}
	\end{tabular}
	\caption{RGB (\sdssi\sdssr\sdssg) color images for four PSZ clusters optically confirmed using our optical imaging. Each panel is centered on the cluster's BCG and has a width of 1 Mpc at the corresponding cluster's redshift. The horizontal bar in the lower right of each panel shows the scale of the panel, where north is up and east is to the left. The location of the PSZ detection is denoted by a red star. The dashed and solid, concentric, red circles are $2'$ and $5'$ in radius respectively.}
	\label{fig:Clusters1}
\end{figure*}

\begin{figure*}
	\centering
	\begin{tabular}{cc}
		\includegraphics[width=0.47\linewidth]{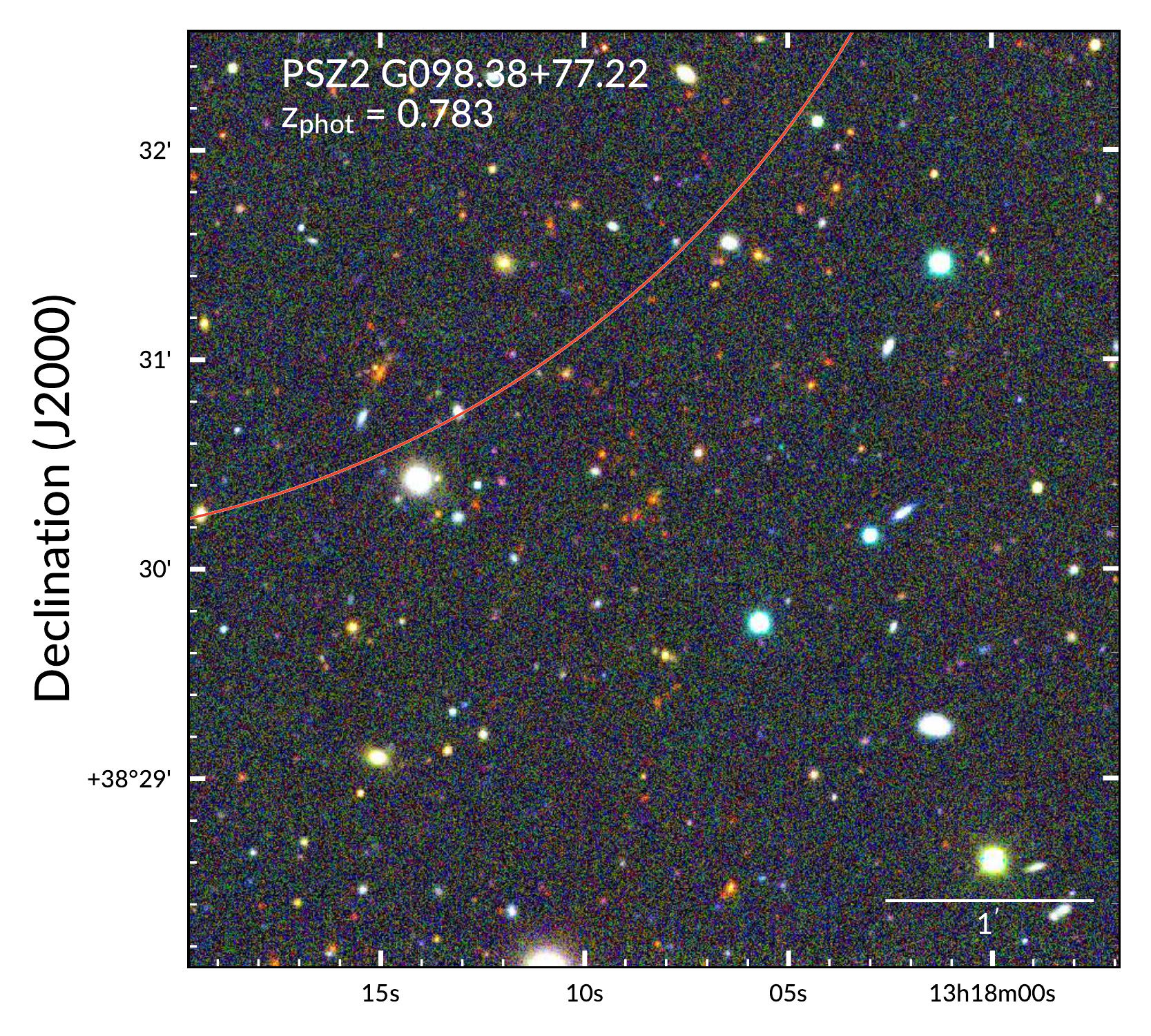}&
		\includegraphics[width=0.47\linewidth]{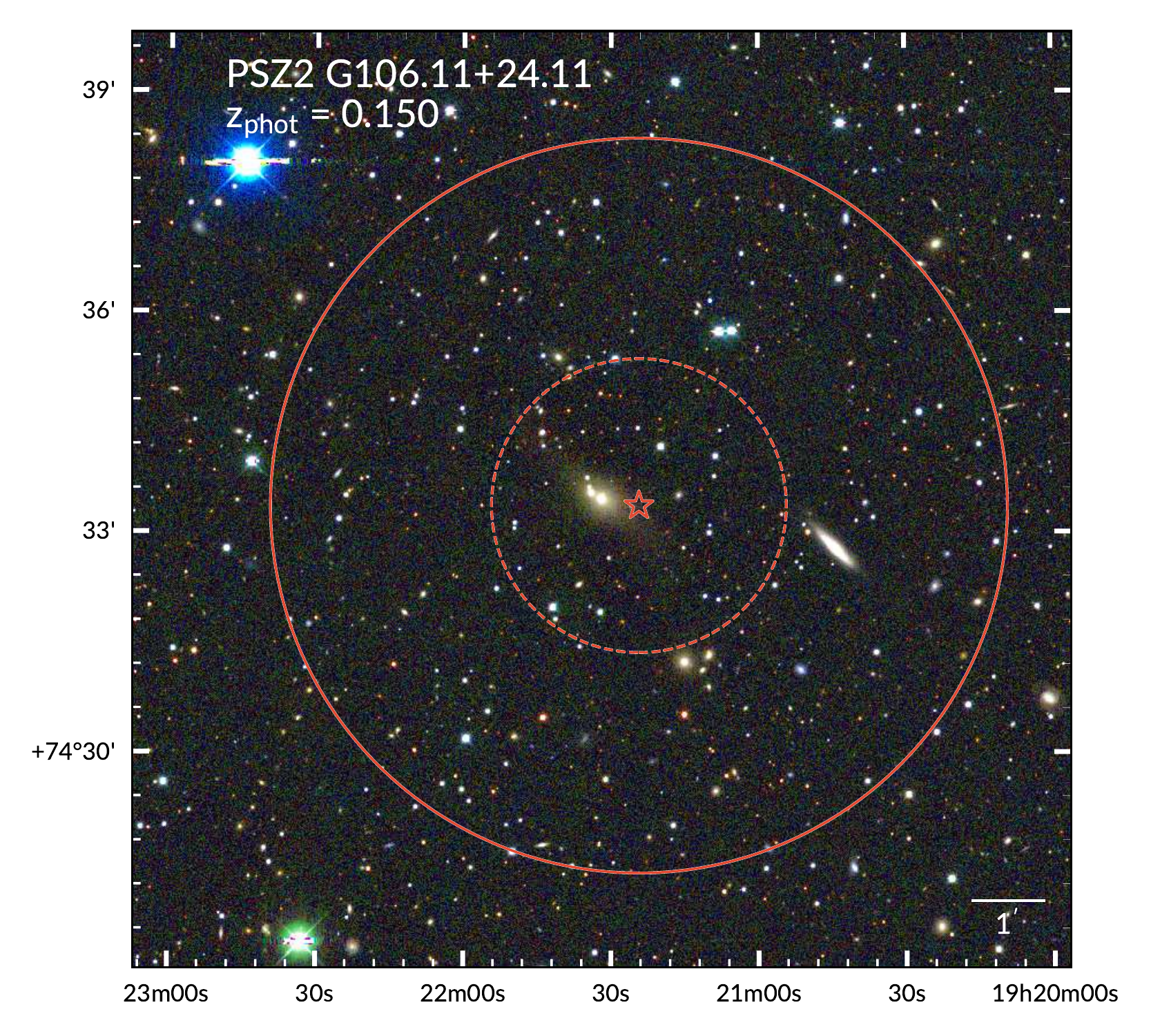}\\
		\includegraphics[width=0.47\linewidth]{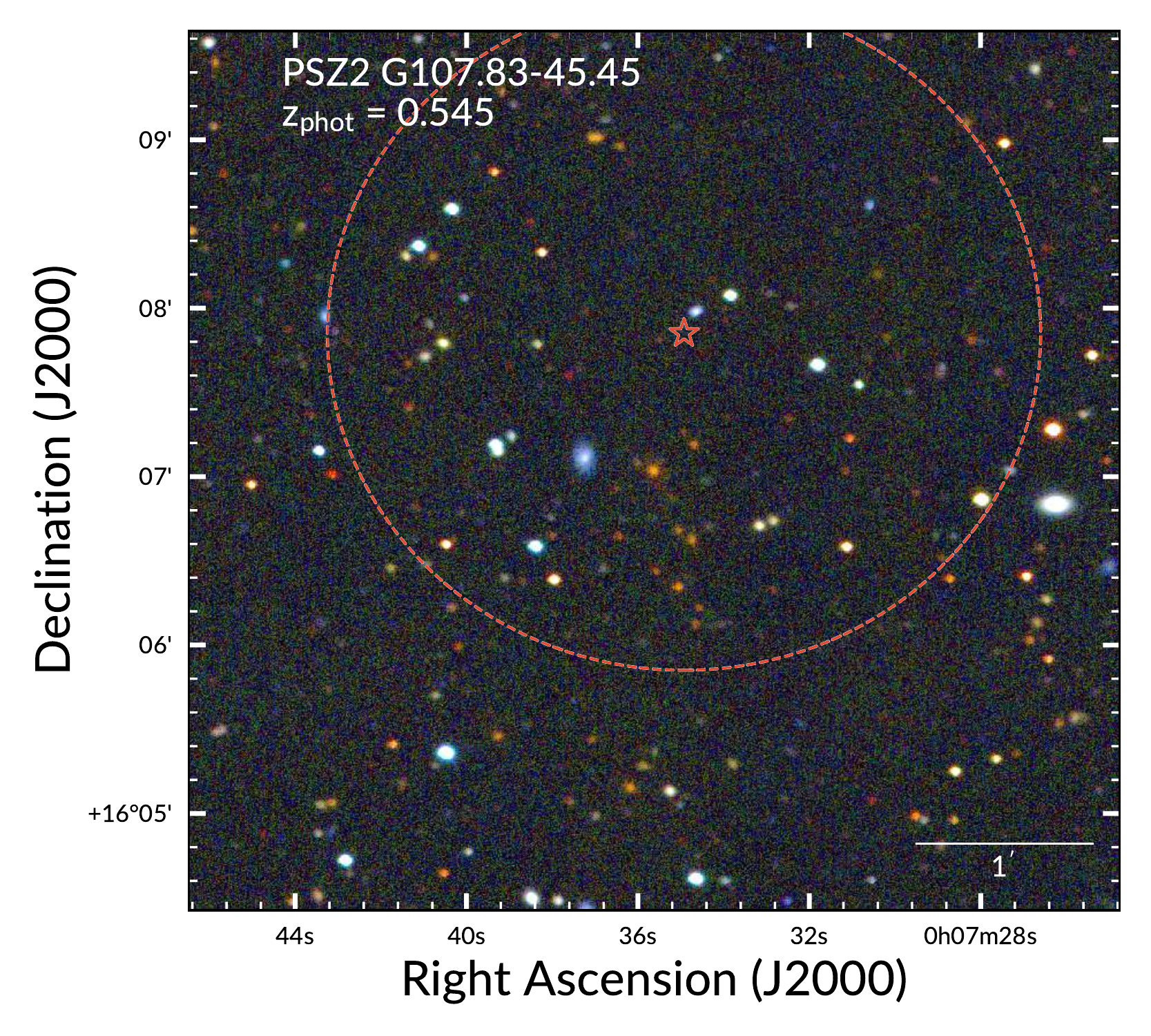}&
		\includegraphics[width=0.47\linewidth]{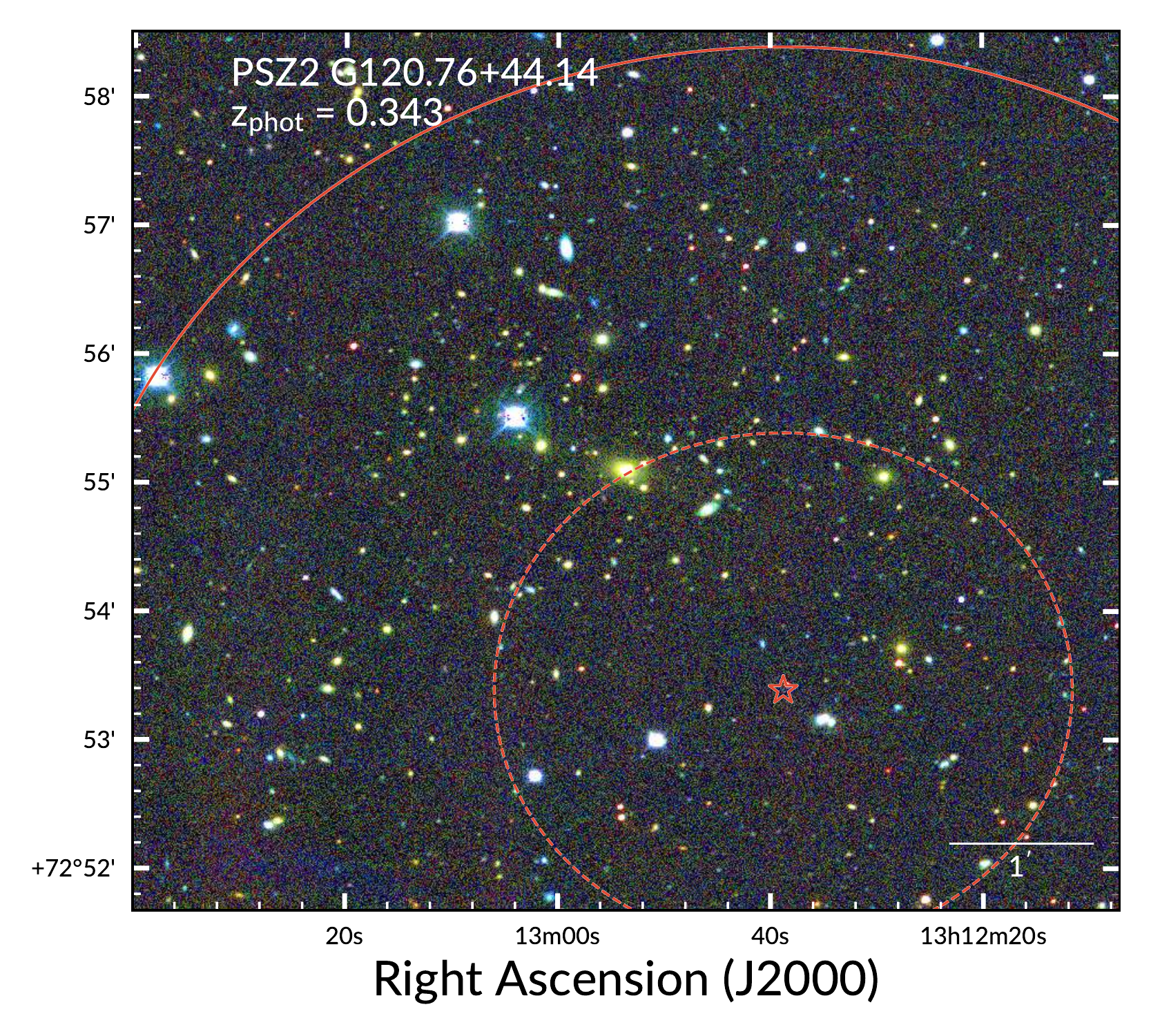}
	\end{tabular}
	\caption{Same as Figure \ref{fig:Clusters1}.}
	\label{fig:Clusters2}
\end{figure*}

\begin{figure*}
	\centering
	\begin{tabular}{cc}
		\includegraphics[width=0.47\linewidth]{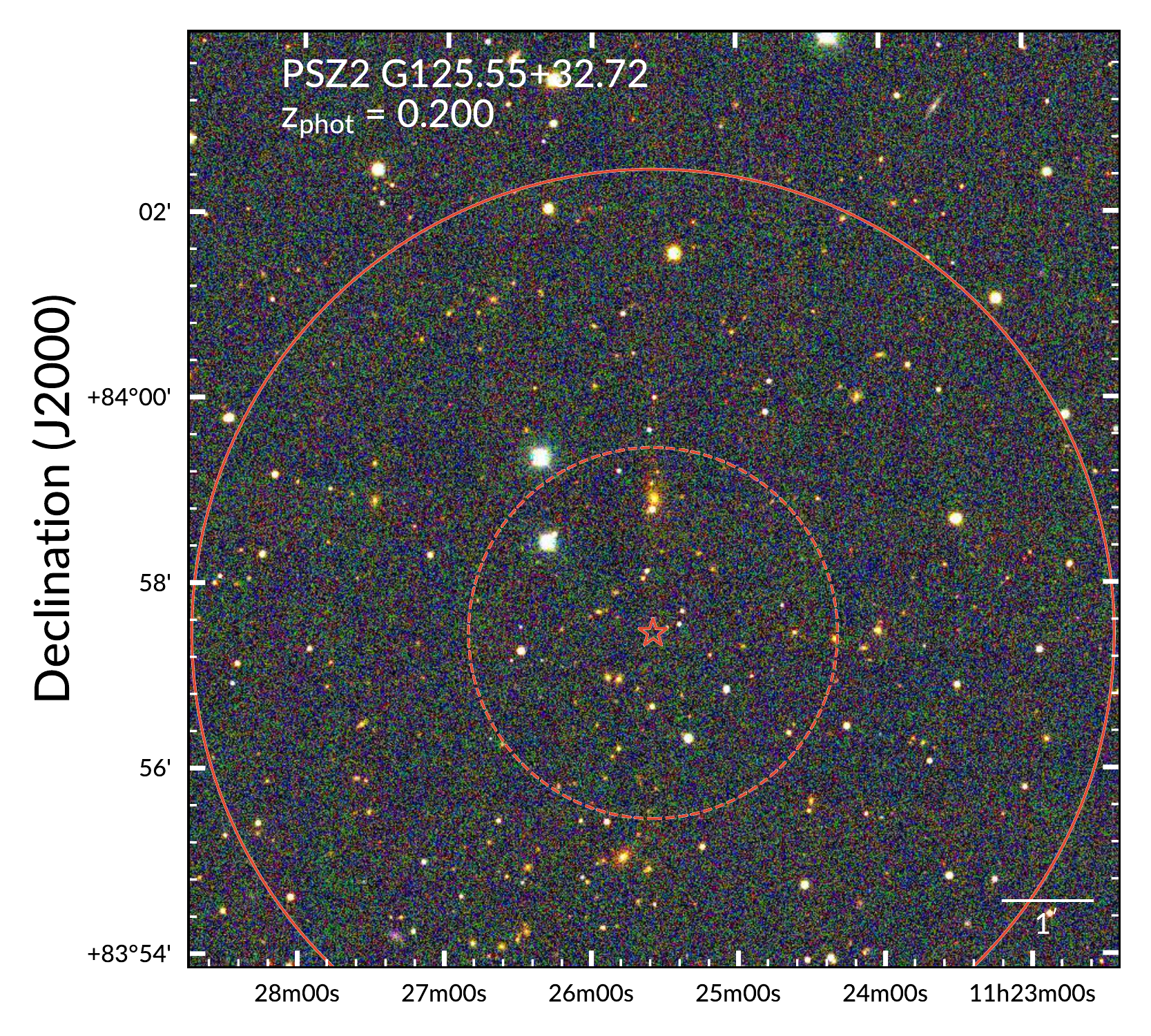}&
		\includegraphics[width=0.47\linewidth]{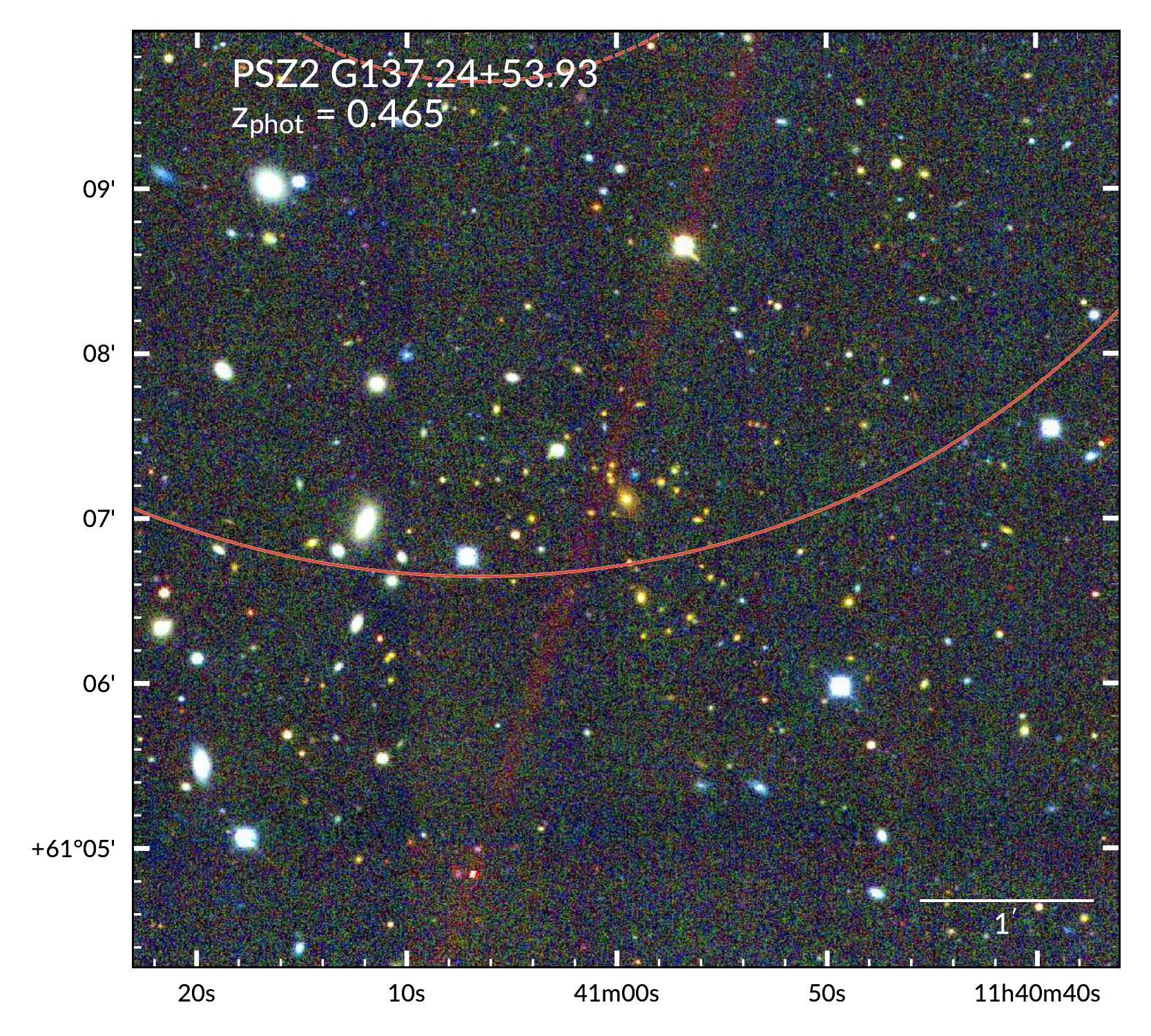}\\
		\includegraphics[width=0.47\linewidth]{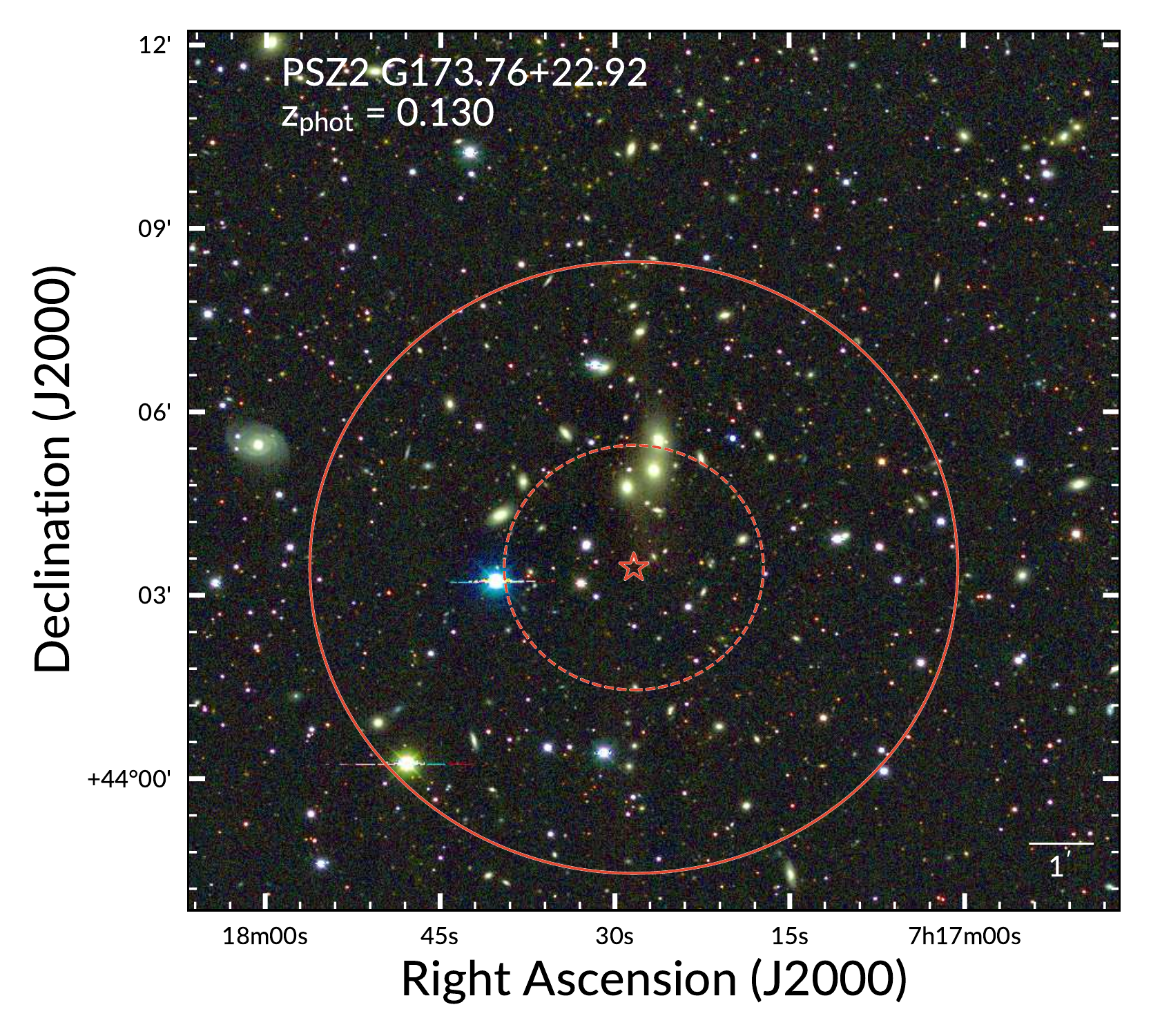}&
		\includegraphics[width=0.47\linewidth]{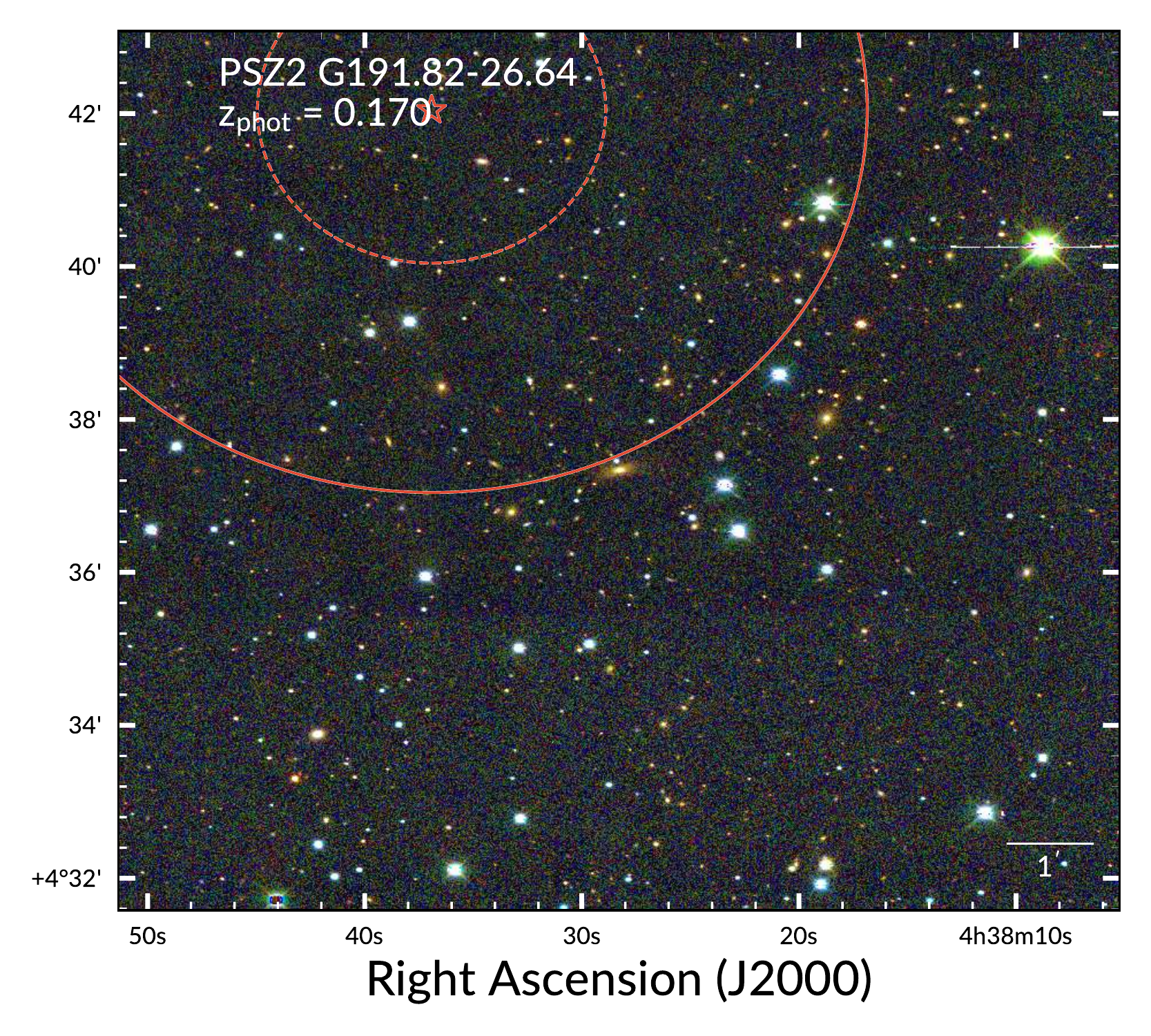}
	\end{tabular}
	\caption{Same as Figure \ref{fig:Clusters1}.}
	\label{fig:Clusters3}
\end{figure*}

\begin{figure*}
	\centering
	\begin{tabular}{cc}
		\includegraphics[width=0.47\linewidth]{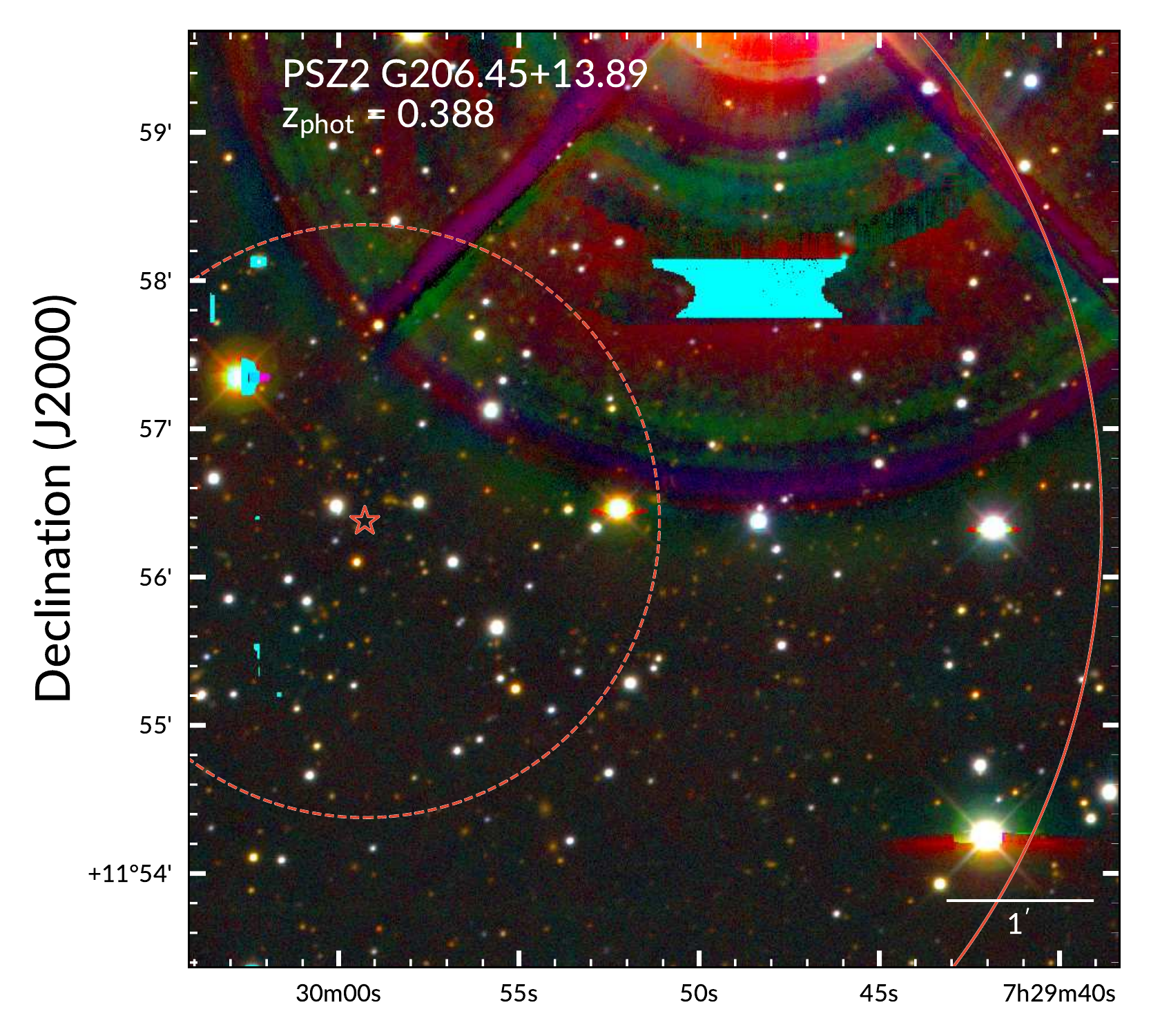}&
		\includegraphics[width=0.47\linewidth]{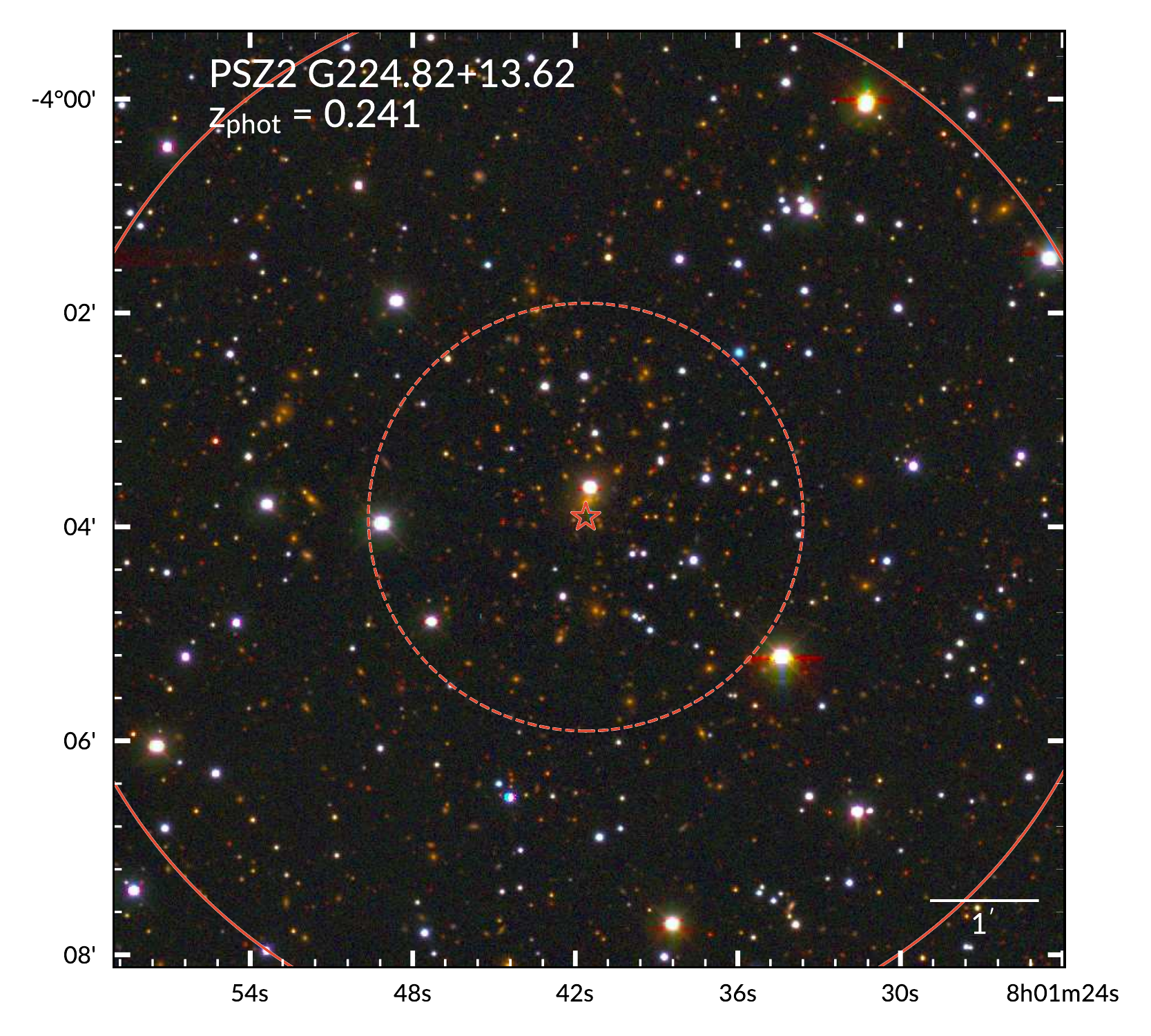}\\
		\includegraphics[width=0.47\linewidth]{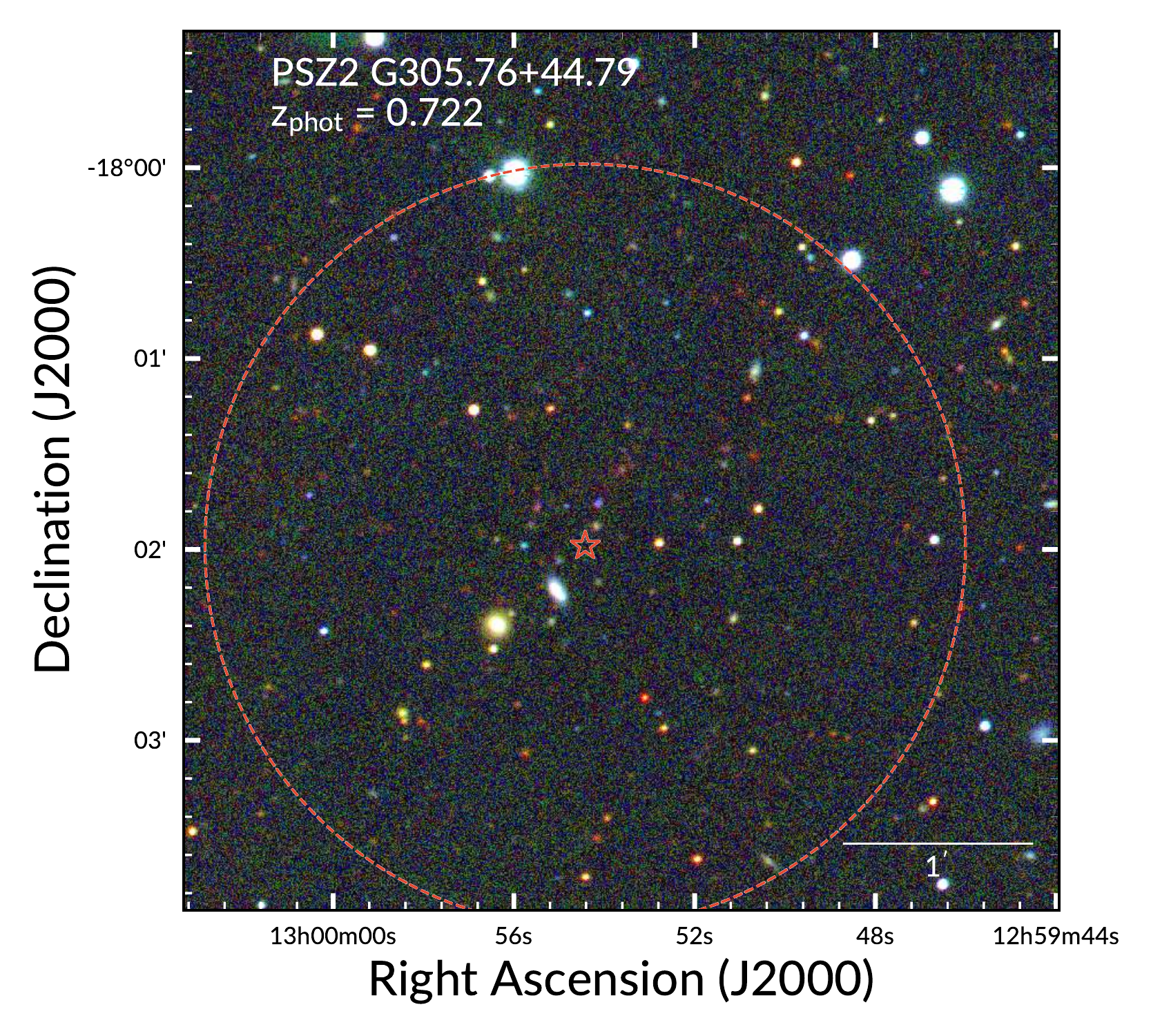}&
	\end{tabular}
	\caption{Same as Figure \ref{fig:Clusters1}.}
	\label{fig:Clusters4}
\end{figure*}

Here, we report the high confidence clusters identified as a result of our cluster finding. A high confidence result consists of a clear BCG and many accompanying satellite galaxies (high richness). For the 85 fields observed, we identify 15 high confidence clusters (see Figures~\ref{fig:Clusters1}--\ref{fig:Clusters4}), including 3 clusters recently confirmed by others. We discuss all 15 clusters in detail below.

In the following subsections, we compare our results with previously known sources by querying the NASA/IPCA Extragalactic Database (NED)\footnote{\url{https://ned.ipac.caltech.edu/}} and the SIMBAD (Set of Identifications, Measurements, and Bibliography for Astronomical Data) astronomical database\footnote{\url{http://simbad.u-strasbg.fr/simbad/}} \citep{Wenger2000}. We include sources from the NRAO (National Radio Astronomy Observatory) VLA (Very Large Array) Sky Survey (NVSS; \citealt{Condon1998}), the R\"{o}ntgensatellit (ROSAT) All-Sky Survey Bright Source Catalog (RASS-BSC; \citealt{Voges1999a}), the ROSAT All-Sky Faint Source Catalog (RASS-FSC; \citealt{Voges2000}), and the SDSS. We have included cross identifications with the all-sky galaxy cluster catalog from \citet{Wen2018} using the label WHY.We make note of confident associations of X-ray and radio sources with the BCG or other clusters members within $5'$ of the reported BCG pointing (within $10'$ for the three low redshift clusters).

\begin{table*}
	\caption[Summary of Cluster Confirmation]{Summary of cluster confirmation: Column 1: The PSZ2 cluster name; Column 2: PSZ2 signal-to-noise ratio; Column 3: PSZ1 ID number; Column 4: BCG Right Ascension in J2000; Column 5: BCG Declination in J2000; Column 6: BCG separation from PSZ position in arcminutes; Column 7: Cluster photometric redshift with 1-$\sigma$ uncertainty; Column 8: Corrected number of member galaxies; Column 9: New confirmation? }
	\begin{threeparttable}
	\centering
	\begin{tabular}{lcccccccc}
	\hline
	Cluster & SNR & PSZ2 (PSZ1) ID & $\alpha$ (J2000) & $\delta$ (J2000) & Sep. ($'$) & $z_\mathrm{cl}$ & Ngal$_c$ & New\\
	(1) & (2) & (3) & (4) & (5) & (6) & (7) & (8) & (9) \\
	\hline
	 PSZ2 G$029.66-47.63$ & 5.74 & 104 & $21:45:29.948$ &  $-21:43:26.35$ & 4.73 & $0.32 \pm 0.03$ & 113 & $\checkmark$ \\
	 PSZ2 G$043.44-41.27$ & 5.55 & 158 & $21:36:43.743$ &  $-10:19:02.08$ & 1.24 & $0.43 \pm 0.03$ & 144 & $\checkmark$ \\
	 PSZ2 G$084.62-15.86$ & 6.01 & 369 (284) & $21:49:42.542$ &  $+33:09:17.54$ & 1.13 & $0.27 \pm 0.10$ & 20 & \\
	 PSZ2 G$096.43-20.89$ & 5.81 & 447 & $22:48:09.417$ &  $+35:33:49.30$ & 0.49 & $0.35 \pm 0.04$ & 76 & $\checkmark$ \\
	 PSZ2 G$098.38+77.22$ & 5.51 & 462 (346) & $13:18:08.288$ &  $+38:30:19.96$ & 5.84 & $0.78 \pm 0.06$ & 50\tnote{a} & $\checkmark$ \\
	 PSZ2 G$106.11+24.11$ & 5.70 & 512 & $19:21:31.903$ &  $+74:33:27.01$ & 0.51 & $0.15 \pm 0.06$ & 27 & $\checkmark$ \\
	 PSZ2 G$107.83-45.45$ & 7.09 & 525 & $00:07:35.617$ &  $+16:07:02.15$ & 0.83 & $0.55 \pm 0.05$ & 29 & $\checkmark$ \\
	 PSZ2 G$120.76+44.14$ & 5.59 & 593 & $13:12:53.600$ &  $+72:55:05.85$ & 2.02 & $0.34 \pm 0.04$ & 81 & $\checkmark$ \\
	 PSZ2 G$125.55+32.72$ & 6.49 & 617 & $11:25:34.186$ &  $+83:58:55.65$ & 1.44 & $0.20 \pm 0.06$ & 44 & $\checkmark$ \\
	 PSZ2 G$137.24+53.93$ & 7.87 & 673 & $11:40:59.546$ &  $+61:07:07.07$ & 4.61 & $0.47 \pm 0.05$ & 42 & $\checkmark$ \\
	 PSZ2 G$173.76+22.92$ & 5.80 & 820 & $07:17:26.646$ &  $+44:05:02.73$ & 1.62 & $0.13 \pm 0.03$ & 117 & $\checkmark$ \\
	 PSZ2 G$191.82-26.64$ & 6.17 & 880 (646) & $04:38:28.285$ &  $+04:37:19.82$ & 5.18 & $0.17 \pm 0.07$ & 29 & $\checkmark$ \\
	 PSZ2 G$206.45+13.89$ & 5.90 & 933 (682) & $07:29:51.241$ &  $+11:56:31.49$ & 1.97 & $0.39 \pm 0.04$ & 72\tnote{a} & \\
	 PSZ2 G$224.82+13.62$ & 5.51 & 1009 (752) & $08:01:41.492$ &  $-04:03:44.48$ & 0.17 & $0.24 \pm 0.04$ & 55 & \\
	 PSZ2 G$305.76+44.79$ & 5.72 & 1441 (1070) & $12:59:53.623$ &  $-18:01:35.22$ & 0.44 & $0.72 \pm 0.07$ & 58\tnote{a} & $\checkmark$ \\
	\hline
	\end{tabular}
	\begin{tablenotes}
		\item[a] Ngal$_c$ should be taken as a lower limit. See Section~\ref{sec:cluster finding} for details.
	\end{tablenotes}
	\end{threeparttable}

\label{tbl:results}
\end{table*}

\subsection{PSZ2 G029.66-47.63} 
This is a rich cluster at $z_\mathrm{cl} = 0.32 \pm 0.03$ with 113 members, approximately $5'$ to the northwest of the \planck\ position. The X-ray source 1RXS J214531.1$-$214339 as well as the cataloged system WHY J214529.9$-$214325 are positionally coincident with the BCG. Our data show another, slightly less rich, system (with $z_\mathrm{cl} = 0.33 \pm 0.04$ and 76 members and with a BCG at $\alpha=21:45:44.7, \delta=-21:47:01.5$) within $0\farcm5$ of the \planck\ position. Both systems likely contribute to the \planck\ SZ signal. This is the richest cluster in our sample.

\subsection{PSZ2 G043.44-41.27} 
This is the richest cluster we have found; the system is at $z_\mathrm{cl} = 0.43 \pm 0.03$ with 144 members. There are two plausible BCGs with nearly the same photo-$z$; the one we select is slightly brighter in the \sdssi-band, yields a slightly higher number of cluster members, and is positionally coincident with the X-ray source 1RXS J$213644.4-101904$. The other bright galaxy is at $\alpha$=21:36:38.6, $\delta$=$-$10:18:35.7, some $1\farcm3$ to the west. This galaxy is associated with a bright radio source (NVSS J213638$-$101836) with a flux density of $107.8 \pm 3.3$ mJy at 1.4 GHz that has been classified as a symmetric double \citep{Douglas1996}.

\subsection{PSZ2 G084.62$-$15.86} 
This cluster was previously confirmed by the \planck\ team, where they quote a spectroscopic cluster redshift (from two members) of $z_\mathrm{spec} = 0.364$ \citep{PlanckCollaboration2016a}. This system has at least three bright member galaxies within $2'$ of the \planck\ position that are plausible BCG candidates. Among these we chose the brightest one in the \sdssi-band (which is about an arcminute south of the \planck\ team's chosen BCG) and recovered the cluster at $z_\mathrm{photo} = 0.27 \pm 0.10$ with 20 members. The \planck\ team's selected BCG is associated with a radio source (NVSS J214940+331031) with a flux density of $19.7\pm 0.8$ mJy at 1.4 GHz.

\subsection{PSZ2 G096.43-20.89} 
The BCG of this cluster is only $0\farcm5$ from the \planck\ position. The cluster's redshift is $z_\mathrm{cl} = 0.35 \pm 0.04$ with 76 members. An X-ray source (1RXS J224806.6+353230) is $1\farcm44$ away from the BCG position. This cluster appears to extend more toward the southwest quadrant in the direction of a cataloged Zwicky cluster (ZwCl $2245.6+3516$; \citealt{Zwicky1968}). Cataloged cluster WHY~J224809.4+353348 is positionally consistent with the BCG of our Planck confirmation.

\subsection{PSZ2 G098.38+77.22} 
This field contains multiple clusters within $10'$ of the Planck candidate. The counterpart we have selected is the one that is closest to the Planck position and also has the highest $N_\mathrm{gal}$ value in our data. It also happens to be the highest redshift cluster in our full sample at $z_\mathrm{cl} = 0.78 \pm 0.06$ with at least 50 members; it is $5\farcm8$ away from the \planck\ position. About $2'$ east of our chosen BCG there is a luminous ($i\sim 20$ mag), red galaxy (SDSS J131814.99+383055.8) with a spectroscopic redshift of $z=0.726$. This galaxy has similar colors to the cluster BCG and may be part of the system. There are two other systems to comment on that are cataloged in NED as GMBCG J199.59552+38.43492 and WHL~J$131842.9+384300$. GMBCG J$199.59552+38.43492$ is a rich system ($N_\mathrm{gal} = 48$) at redshift $z_\mathrm{cl}=0.41 \pm 0.05$) that is located some $9'$ south of the Planck position.
\hbox{WHL J$131842.9+384300$} is not as rich ($N_\mathrm{gal} = 23$) and not as distant ($z_\mathrm{cl}=0.17 \pm 0.05$), and is $8\farcm6$ north of the Planck position. Each of the BCGs in these clusters have spectroscopic redshifts (0.4193 and 0.2377) that are consistent with the values quoted from our photometry. It is possible that all of the clusters discussed here contribute to the reported SZ signal.

\subsection{PSZ2 G106.11+24.11} 
PSZ2 G106.11+24.11 is a low redshift system at $z_\mathrm{cl} = 0.15 \pm 0.06$. The BCG is a large galaxy close to the \planck\ position. There are 27 members after correcting for background galaxies. This object was identified as an X-ray cluster (RXC J$1921.3+7433$) based on its extent in the RASS \citep{Bohringer2000}, although these authors did not publish a redshift. Using their flux value and our photometric redshift, we estimate the cluster's X-ray luminosity in the 0.1-2.4 keV band to be $L_\mathrm{X} \sim 2.6\times 10^{44}$ erg s$^{-1}$.
This luminosity value is broadly consistent with those of other confirmed \planck\ clusters at this redshift range \citep{PlanckCollaboration2015}. We note that the cataloged cluster WHY J192115.0+743114 is centered on a fainter member of our system some $2'$ south of the Planck position.

\subsection{PSZ2 G107.83-45.45} 
This cluster at $z_\mathrm{cl} = 0.55 \pm 0.05$ has 29 members. The BCG (SDSS J$000735.62+160701.8$) as well as another member (SDSS J$000736.15+160508.9$) have spectroscopic redshifts from the SDSS of $z=0.5673$ and $z=0.5667$, respectively. There are 4 other galaxies with SDSS spectroscopic redshifts (0.5649, 0.5661, 0.5655, and 0.5625) within $7'$ (2.7 Mpc) of the BCG.

\subsection{PSZ2 G120.76+44.14} 
The BCG we select yields an estimated redshift for the cluster of $z_\mathrm{cl} = 0.34 \pm 0.04$ with 81 members; this galaxy has a published spectroscopic redshift of $z=0.2959$ \citep{Huchra1990}. We associate the cluster with Abell 1705 and the RASS X-ray source 1RXS J$131252.0+725514$.

Approximately $2\farcm6$ south of the PSZ2 position, there is a luminous red galaxy with a gravitationally lensed arc that is the BCG of a rich system cataloged as WL $1312.5+7252$ with a photometric redshift of 0.55 \citep{Dahle2003}. This galaxy is also a radio source (NVSS J$131230+725051$) with a flux density of $3.5 \pm 0.5$ mJy (1.4 GHz). Our analysis also yields a rich cluster with this BCG ($\alpha$=13:12:30.9, $\delta$=+72:50:54.2) with Ngal $= 61$ at $z=0.60 \pm 0.05$. Given the similar optical richnesses of these two systems it is likely that both contribute to the \planck\ SZ signal.


\subsection{PSZ2 G125.55+32.72} 
There are two plausible BCGs in this cluster; the one we select yields a redshift of $z_\mathrm{cl} = 0.20 \pm 0.06$ and 44 members. The other BCG ($\alpha$=11:25:46.8, $\delta$=+83:55:04.4) is about 0.12 mag fainter in the \sdssi-band and, using it for cluster finding, results on a cluster at $z_\mathrm{cl} = 0.20 \pm 0.05$ with 46 members. These two plausible BCGs are separated by $\sim4'$. Both BCGs are radio sources: the northern galaxy (corresponding to our selected BCG) is cataloged as NVSS J$112535+835858$ with a 1.4 GHz flux density of $10.2 \pm 0.9$ mJy; the southern one corresponds to NVSS J$112550+835508$ with a flux density at the same frequency of $3.6 \pm 0.6$ mJy. The southern galaxy is about $1'$ away from the RASS X-ray source 1RXS J$112547.3+835559$. Both systems should contribute to the \planck\ SZ signal and the quoted richness in Table~\ref{tbl:results} is almost surely an underestimate of the richness of the combined system.


\subsection{PSZ2 G137.24+53.93} 
Here we find a cluster at $z_\mathrm{cl} = 0.47 \pm 0.05$ with 42 members. The BCG is a radio source (NVSS J$114059+610658$) with a flux density of $10.2 \pm 0.9$ mJy (1.4 GHz) and has a spectroscopic redshift from the SDSS of $z=0.4770$. Another cluster member also has a concordant SDSS spectroscopic redshift of $z=0.4697$. The cluster has been cataloged as WHL J$114058.8+610631$ with a similar photometric redshift. We note that the cataloged cluster GMBCG J$175.54011+61.13607$ is some $8\farcm5$ east of the Planck position. This is a low redshift system ($z_\mathrm{cl} = 0.13 \pm 0.06$) with 20 members. It is disfavored as the counterpart given its large distance from the Planck position.

\subsection{PSZ2 G173.76+22.92} 
This low redshift system, which we find at $z_\mathrm{cl} = 0.13 \pm 0.03$, has a very interesting BCG. It is cataloged in NED as \hbox{B3 $0713+441$} and has a spectroscopic redshift of $z=0.0652$ \citep{Bauer2000}. It is also associated with the RASS X-ray source 1RXS J071726.9+440557 as well as a bright radio source (NVSS J$071726+440504$) with a flux density of $220.4\pm 7.6$ mJy (at 1.4 GHz). Higher resolution images from Faint Images of the Radio Sky at Twenty-centimeters (FIRST; \citealt{Becker1995}) reveal that this radio source is double lobed. Some 6$^\prime$ to the east is a cataloged Seyfert 1 galaxy (2MASX J$07180060+4405271$) with a spectroscopic redshift of $z=0.0614$
\citep{Michel1988} and a 1.4 GHz radio flux of $50.8 \pm 1.6 $ mJy (NVSS J$071800+440527$). Our cluster is positionally coincident with  WHY~J071726.7+440502.


\subsection{PSZ2 G191.82-26.64} 
This is another low redshift cluster at $z_\mathrm{cl} = 0.17 \pm 0.07$ with 29 members. Two cluster members are associated with radio sources: NVSS J$043836+043824$ and NVSS J$043818+043802$ with 1.4 GHz flux densities of $24.8 \pm 1.2$ mJy and $22.7 \pm 1.5$ mJy, respectively.

\subsection{PSZ2 G206.45+13.89} 
We find a cluster at $z_\mathrm{cl} = 0.39 \pm 0.04$ with at least 72 cluster members. A bright star (V $= 4.5$ mag; \citealt{Hog2000}) lies only $\sim4\farcm9$ away from the reported BCG, which prevents accurate photo-$z$ estimates for a significant fraction of the projected area of the cluster. This cluster has been previously confirmed in \cite{Barrena2018} as a rich cluster with a spectroscopic redshift of $z_\mathrm{spec} = 0.406$ from 45 members. We confirm the presence of a possible gravitationally-lensed arc $\sim13''$ northeast of the BCG.

\subsection{PSZ2 G224.82+13.62} 
The BCG of this system is partially obscured by a nearby star and was not fully deblended in our catalogs. Still we are able to find a rich cluster at $z_\mathrm{cl} = 0.24 \pm 0.04$ with 55 members. An interesting aspect of this cluster is that it is positionally coincident with an unidentified X-ray source (2E $0759.2-0355$) from the \textit{Einstein Observatory} \citep{Harris1990}. This cluster was confirmed by \cite{Barrena2018} with a spectroscopic redshift of $z_\mathrm{spec} = 0.274$ from 28 members.

\subsection{PSZ2 G305.76+44.79} 
Finally, PSZ2 G$305.76+44.79$ is our second highest redshift cluster at $z_\mathrm{cl} = 0.72 \pm 0.07$ with at least 58 members. The BCG is associated with radio source PMN J$1259-1801$ with a $1.4$ GHz flux density of $42.2\pm 1.4$ mJy from the NVSS.


\section{Discussion}\label{sec:discussion}
In this section, we discuss the results given in the previous section as a whole, and frame those results in the context of the broader PSZ sample.

\subsection{Overview of Sample}
Of the 85 cluster candidates observed as part of our program, we confirm 15 clusters, 12 of which were previously unrecognized as associations with \planck\ candidates. The SNR range of our observed sample covers values 5.06 to 15.89 with a median value of 6.35. The sample of confirmed clusters has a considerably lower range of SNR values ($5.51 - 7.87$) with a lower median of 5.80. The clusters span the redshift range $ 0.13 < z < 0.78$ with mean redshift $<z> = 0.36$. Seven of the clusters have published spectroscopic redshifts from other observations (see Section~\ref{sec:results} for details). The scatter between our reported photo-$z$'s and the reported spec-$z$'s, using Equation~\ref{eqn:scatter}, is $\sigma_f = 0.036$ indicating that our clusters' photometric redshifts are accurate.

As part of the confirmation process we limited our cluster search to objects within $6'$ of the reported PSZ position. The mean separation between our reported BCGs and the PSZ position is $2\farcm15$ with 68\% of BCGs within $2'$ of the quoted PSZ position. We find that our spatial distribution of clusters is roughly consistent with other follow up observations of PSZ cluster candidates \citeeg{Barrena2018}.

A number of clusters have noteworthy properties. We find that roughly $1/3$ of the confirmed sample appear to contain multiple BCGs, or are part of multi-cluster systems. Seven of the 15 clusters have BCGs or other cluster members that host radio sources.

\subsection{Implications for Full PSZ Sample}
Using our galaxy cluster identification scheme described in Section~\ref{sec:cluster finding}, we fail to identify an optical counterpart to 70 PSZ cluster candidates. Because our observations are limited to candidates with SNR $>5$-$\sigma$, we would expect at most one failed detection. Here we discuss three possible scenarios that could contribute to this low purity.

\begin{figure}
	\includegraphics[width=\columnwidth]{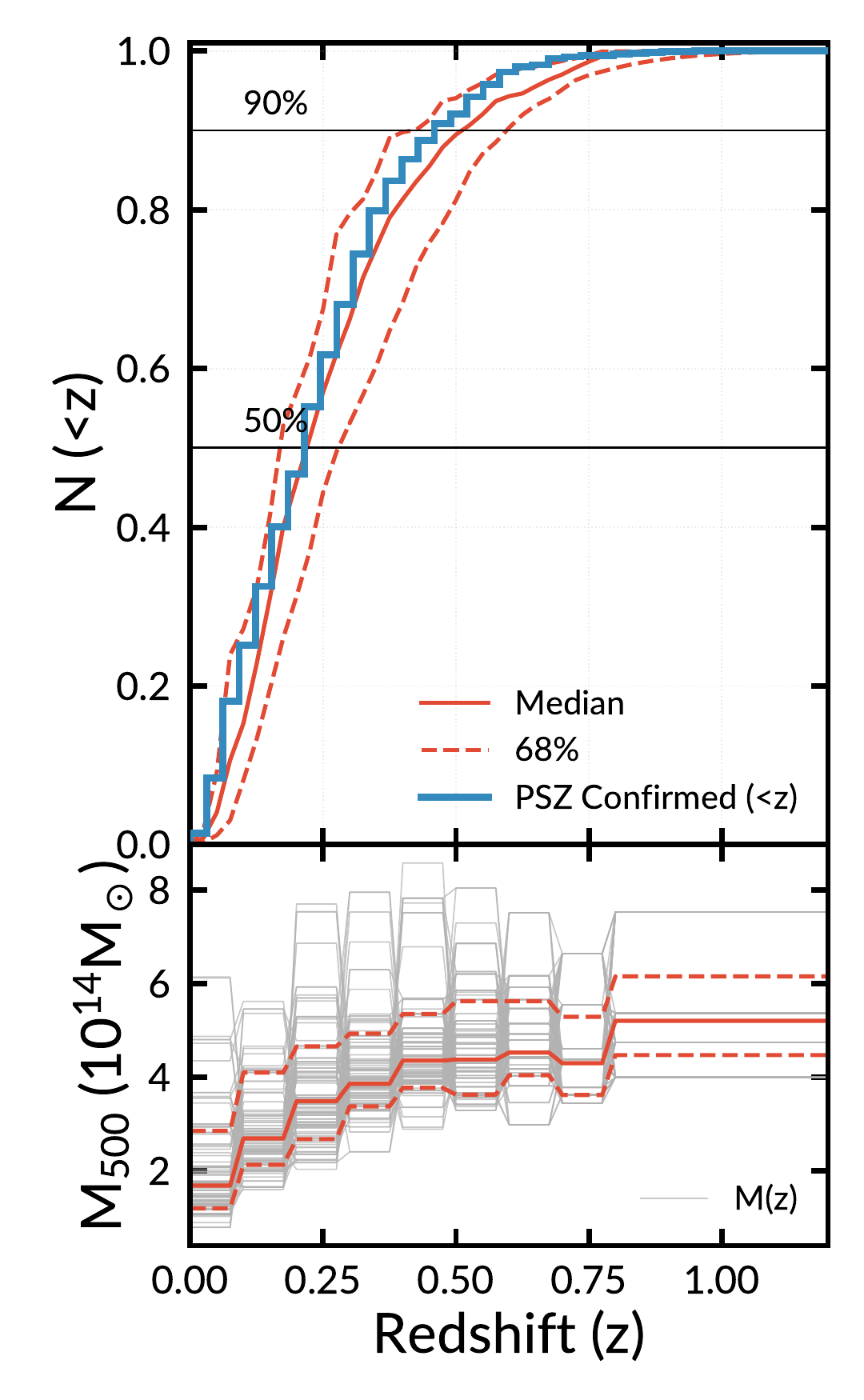}
	\caption{Top: The predicted number of clusters as a function of redshift and normalized to unity from the mass function of \cite{Tinker2008}. At each redshift the lower mass limit is given by the mass in the bottom panel at the corresponding redshift. The blue PSZ Confirmed curve shows the normalized, cumulative number of confirmed clusters as a function of redshift. For reference we show the 50\% and 90\% fractional completeness levels. Bottom: Cluster mass as a function of redshift. Redshift bins are the same as \cite{PlanckCollaboration2015a} Figure 27. In each panel the solid and dashed orange lines show the median and region enclosing 68\% of the data respectively.}
	\label{fig:cluster_forecast}
\end{figure}

The first possibility is that clusters in our sample were missed because they are at low-$z$. The initial design of our survey restricts the cluster search to a $6'$, radius, search window centered on the PSZ position. Low-$z$ clusters could appear as an isolated elliptical galaxy or a small number of galaxies and thus might not be classified as clusters. In follow up inspection of the full--sized mosaics, approximately 1 \degsq, reveals no such low-$z$ structures.

To investigate this low-$z$ hypothesis quantitatively, we compute the expected fraction of galaxy clusters as a function of redshift from the cluster mass function of \cite{Tinker2008}. The cluster mass sensitivity of \planck\ varies with redshift (see Figure 27, \citealt{PlanckCollaboration2015a}). To capture this variability, we draw random samples (with replacement) from the confirmed sample of clusters as reported in \cite{PlanckCollaboration2015a}. Because each cluster is previously confirmed, the PSZ2 catalog provides an M$_{500c}$ estimate of the total mass. The bottom panel of Figure~\ref{fig:cluster_forecast} shows the cluster mass as a function of redshift, where the gray lines are individual samples and the orange solid and dashed curves are the median and 68\% limits respectively. The top panel of Figure~\ref{fig:cluster_forecast} shows the expected cumulative distribution of galaxy clusters as a function of redshift.

From these numerical simulations we can see that a low fraction ($<5\%$) of all clusters detectable by \planck\ are expected to reside at redshifts less than $0.1$. Of the 85 fields observed, we expect approximately four clusters to lie at such low redshifts. This is consistent with the three low-$z$ clusters we recover during the course of our analysis. Thus it is very unlikely that we have missed a significant number of low redshift clusters.

The second possibility posits that many  of the clusters in our sample lie at redshifts beyond our optical detection limits. The median limiting \sdssi-band magnitude of our survey is $23.2$ mag, corresponding to a limiting redshift of $z=0.75$ for an \lstar\ galaxy (see Figure~\ref{fig:recovery_redshift}).

From the predicted number of PSZ clusters as a function of redshift (Fig.~\ref{fig:cluster_forecast}), we expect $\sim3\%$ of the sample to lie at $z>0.7$. Out of the 1943 total \planck\ SZ detections, our simulated redshift distribution predicts that there should be 58 clusters at high-$z$ (if all \planck\ SZ detections are real clusters).  There are only 12, currently confirmed, clusters at $z>0.7$ in PSZ2, only five of which are above $z=0.8$. So there should be 46 high-$z$ clusters among the unconfirmed sample of 613, or approximately 7\%. Applying this percentage to the 85 fields we observed, suggests that we should have found six high-$z$ clusters. We have only found two, each of which lies close to the redshift limit of our optical search.

Thus while high-$z$ systems were expected to be missed by this (optical only) survey, high-$z$ clusters alone cannot account for the high number of still unconfirmed clusters. We expect further follow up observations with deep infrared imaging will be required to either confirm or place additional limits on putative high-$z$ clusters.

\begin{figure}
	\includegraphics[width=\columnwidth]{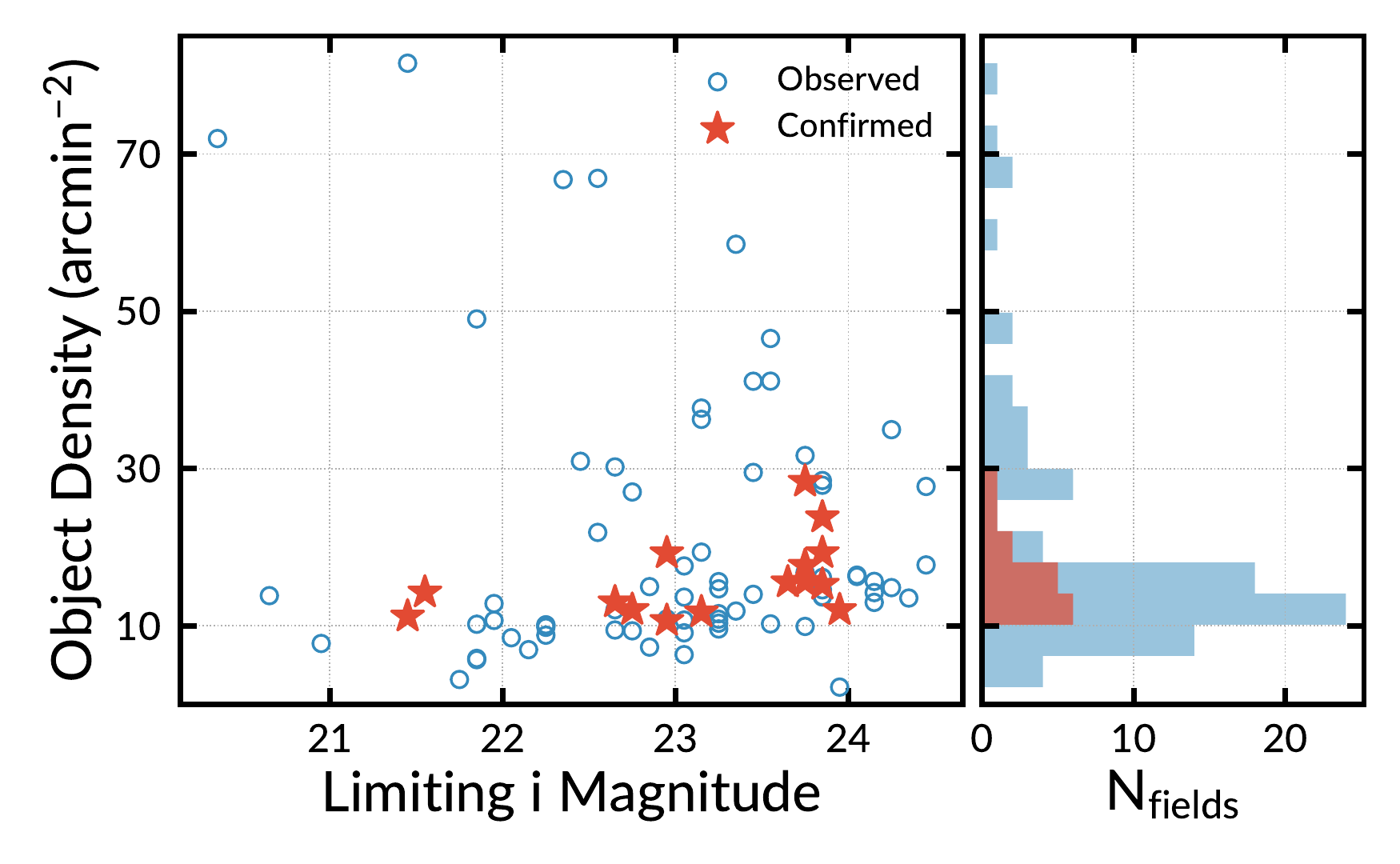}
	\caption{Right: Object Density in units of number per \arcminsq as a function of the 80\% \sdssi-band limiting magnitude taken from Figure~\ref{fig:recovery_redshift}. Fields where no cluster is identified are given by blue circles and fields with confirmed clusters are shown by orange stars. Left: Histogram of the object surface density of observed fields.}
	\label{fig:N_vs_density}
\end{figure}

The third possibility is that the underlying cause for failed cluster identification is confusion from source crowding in our images. For example, if a cluster lies behind a dense foreground of unrelated sources, then it could be difficult to visually identify the BCG and the corresponding fainter, member galaxies. To estimate the object surface density in our fields, we sum the number of objects reported by \textsc{SExtractor} and divide by the sky-area of the image. Figure~\ref{fig:N_vs_density} allows us to explore the possibility that we fail to confirm a cluster in the majority of fields due to unrelated sources. The left panel shows the surface density of objects in the search area as a function of the limiting \sdssi-band magnitude (see also the left panel of Figure~\ref{fig:recovery_redshift}). The right panel shows the number of observed fields with the corresponding object density. In both panels, the fields where we do not identify a cluster are shown in blue, and the fields with successfully identified clusters are shown in orange.

Our identified clusters fall in a relatively narrow range of object densities, $10-30$ objects \perarcminsq. Roughly 2/3 of our fields fall within this range. Of the 31 fields outside the range, 15 fields have higher densities. If all 15 fields contain an unidentified cluster we could still not account for the vast majority of fields where we did not identify a cluster, 51 fields or $\sim60\%$ of our observed sample.

Perhaps more interesting are the 16 fields where the object surface density is lower than the range where we made successful identifications. These are fields where it should be relatively easy to identify a massive cluster should it exist. Of course, the presence of a massive cluster could raise the object density above the lower identification threshold. These fields, especially the fields with deep optical limiting magnitudes are prime candidates for the potential high-$z$ clusters lurking beyond the reach of our optical search.

An alternative to the above possibilities is that the vast majority of cluster candidates in our sample are not true 5-$\sigma$ (or higher) detections.	In other words, no optical counterpart to the PSZ detection exists. One possibility for the low confirmation fraction is contamination of the Planck SZ signal by unrelated radio source emission. To investigate this, as part of our NED search, we track NVSS radio sources within $5'$ of the reported PSZ position. We find approximately 75\% of confirmed PSZ clusters have a NVSS radio source (39.6 mJy average flux) and unconfirmed PSZ candidates show slightly fewer sources with approximately 55\% having a NVSS radio source (25.1 mJy average flux) within $5'$. For the 85 cluster candidates observed in this work $\sim70\%$ of fields have at least one radio source (30.8 mJy average flux). Twelve of the 15 confirmed clusters (80\%; 26.12 mJy average flux) have nearby radio sources. Based on these results, we find it unlikely that radio contamination is a leading cause of the low purity of the unconfirmed PSZ sample.

The search method we employed for this study has focused on identifying optically rich systems. This clearly assumes that the unconfirmed clusters are like previously confirmed clusters from \planck, SPT, and ACT. Although this is a fully justified assumption, it also means that we have potentially missed identifying a class of optically poorer clusters associated with the \planck\ candidates. Indeed the low identification fraction in our work suggests that many aspects of the confirmation process (optical band searches, restricting counterparts to a specified separation from the cataloged \planck\ position, requiring high optical richness for the counterparts, and so on) may need to be re-evaluated.

\section{Summary}\label{sec:summary}
In this work, we report on our analysis of images targeting galaxy cluster candidates selected from the second \planck\ all-sky galaxy cluster catalog (PSZ2) \citep{PlanckCollaboration2015a}. We observe 85 candidates with SZ effect SNR $>5$ over the course of 17 nights spread over three years (2014--2017). We independently develop and utilize a pipeline to process the \sdssg\sdssr\sdssi\sdssz\ imaging taken with both the MOSAIC 1.1 and MOSAIC 3 imagers on the KPNO Mayall-4m telescope (see Section~\ref{sec:data reduction}). We present the first results from the complete data set of 85 fields, 15 rich galaxy clusters of which 12 were previously unassociated with a \planck\ cluster.

After computing accurate photometric redshifts (see Section~\ref{sec:analysis}) we visually inspect each galaxy cluster candidate and adopt stringent confirmation criteria based on the richness and separation from the reported PSZ2 position.

The newly discovered clusters range in photometric redshift between $0.13 < z < 0.78$. The upper redshift limit is due to the depth of imaging restricting our ability to reliably detect \lstar\ galaxies to $z<0.9$. Ultimately, this prevents us from finding the most interesting rich clusters at very high redshifts; this will be addressed in a future follow-up study.

A large motivation for this work has come from the recent successes of other SZ follow-up programs (see citations in Section~\ref{sec:intro}). We present this sample of clusters to aid in the confirmation of the full PSZ cluster sample, and to potentially reveal clusters with interesting astrophysical properties. The PSZ cluster candidates which remain unconfirmed have the potential to be the most interesting.

In future work, we will continue to re-evaluate our confirmation methodology to explore the possibility that other lower, richness systems are potential counter parts to the \planck\ sample. Additionally, we expect a few higher redshift clusters in the sample of 85 included in this survey, and will explore this with multi-wavelength studies.

All source code used to conduct this analysis is available at \url{https://github.com/boada/planckClusters}. The raw and processed imaging along with final data products for the 15 clusters presented here are available upon request.

\section*{Acknowledgments}
This work was supported by NASA Astrophysics Data Analysis grant number NNX14AF73G and NSF Astronomy and Astrophysics Research Program award number 1615657. LFB acknowledges support from CONICYT project Basal AFB-170002. We thank the several undergraduate students who worked on this project: August Miller, Alexis Freiling, Jessica Kerman, Kevin Feigelis, Ingrid Zimmerman, and Cole Bisaccia. Alexis, Jessica, and Ingrid were participants in Project SUPER (Science for Undergraduates: A Program for Excellence in Research), a STEM-focused enrichment program through Douglass that offers undergraduate women the opportunity to actively participate in academic research. August participated in the 2014 Rutgers NSF REU program.
This research made use of several open source packages: \textsc{APLpy}\footnote{\url{http://aplpy.github.com}}, an open-source plotting package for Python; the \textsc{IPython}\footnote{\url{https://ipython.org}} package \citep{Perez2007}; \textsc{matplotlib}\footnote{\url{https://matplotlib.org}}, a Python library for publication quality graphics \citep{Hunter2007} and \textsc{Astropy}\footnote{\url{http://www.astropy.org}}, a community developed core Python package for Astronomy \citep{TheAstropyCollaboration2013}.
Funding for the SDSS and SDSS-II has been provided by the Alfred P. Sloan Foundation, the Participating Institutions, the National Science Foundation, the U.S. Department of Energy, the National Aeronautics and Space Administration, the Japanese Monbukagakusho, the Max Planck Society, and the Higher Education Funding Council for England. The SDSS is managed by the Astrophysical Research Consortium for the Participating Institutions.
This work has made use of data from the European Space Agency (ESA) mission \textit{Gaia}\footnote{\url{https://www.cosmos.esa.int/gaia}}, processed by the \textit{Gaia} Data Processing and Analysis Consortium\footnote{\url{https://www.cosmos.esa.int/web/gaia/dpac/consortium}} (DPAC). Funding for the DPAC has been provided by national institutions, in particular the institutions participating in the \textit{Gaia} Multilateral Agreement.
This research has made use of the VizieR catalogue access tool, CDS, Strasbourg, France. The original description of the VizieR service was published in \cite{Ochsenbein2000}.
This research has made use of the SVO Filter Profile Service\footnote{\url{http://svo2.cab.inta-csic.es/theory/fps/}} supported from the Spanish MINECO through grant AyA2014-55216.
This research has made use of the NASA/IPAC Extragalactic Database (NED) which is operated by the Jet Propulsion Laboratory, California Institute of Technology, under contract with the National Aeronautics and Space Administration.
Finally, we thank all the staff at KPNO for ensuring the success of our observing and aiding in the acquisition of our data.

\bibliographystyle{apj}

\end{document}